\documentclass[subeqn,11pt, a4paper]{amsart}

\usepackage[margin=35mm]{geometry}

\usepackage{amssymb,amsmath,amsfonts,amsthm, amscd, mathrsfs, helvet, mathtools}
\usepackage{bm, stmaryrd}
\usepackage{graphicx, verbatim}
\usepackage[mathcal]{euscript}
\usepackage{color}
\usepackage[all]{xy}
\usepackage{relsize}

\usepackage{tikz}
\usetikzlibrary{arrows,matrix,calc}
\usetikzlibrary{decorations.markings,shapes.geometric,shapes.misc}
\usetikzlibrary{decorations.pathreplacing,positioning}  

\tikzset{cross/.style={cross out, draw=black, minimum size=2*(#1-\pgflinewidth), inner sep=0pt, outer sep=0pt}, cross/.default={1pt}}

\allowdisplaybreaks[1]

\usepackage{pict2e}



\definecolor{myPurple}{rgb}{0.5,0.1,0.6}
\definecolor{myOrange}{rgb}{1.0,0.5,0.0}
\definecolor{myRed}{rgb}{1.0,0.0,0.0}
\definecolor{myGreen}{rgb}{0.0,0.5,0.0}
\definecolor{LatexBlue}{rgb}{0.211765,0.227451,0.666667}
\definecolor{myBlue}{rgb}{0.0,0.0,1.0}
\definecolor{myBlack}{rgb}{0.0,0.0,0.0}
\definecolor{myGray}{rgb}{0.3,0.3,0.3}

\theoremstyle{plain}
\newtheorem{theorem}{Theorem}[section]
\newtheorem*{theorem*}{Theorem}
\newtheorem{proposition}[theorem]{Proposition}
\newtheorem*{proposition*}{Proposition}

\newtheorem{lemma}[theorem]{Lemma}

\theoremstyle{remark}

\newtheorem{remarkx}{Remark}[section]

\newenvironment{remark}
  {\pushQED{\qed}\remarkx}
  {\popQED\endremarkx}

\newcommand{\tensor}[1]{{\mathfrak{#1}}}

\DeclareMathOperator{\res}{res}

\DeclareMathOperator{\End}{End}

\DeclareMathOperator{\Ad}{Ad}

\def\d{\mathfrak{d}}
\def\g{\mathfrak{g}}

\def\id{\textup{id}}

\def\k{\mathfrak{k}}

\def\p{\mathfrak{p}}

\def\t{\mathfrak{t}}

\def\L{\mathcal{L}}

\def\ha{\mbox{\small $\frac{1}{2}$}}

\newcommand{\lau}[1]{(\kern-.2em( #1 )\kern-.2em)}

\newcommand{\ms}[1]{\mathsf{#1}}

\def\CC{\mathbb{C}}
\def\CP{\mathbb{C}P^1}

\def\RR{\mathbb{R}}

\def\ZZ{\mathbb{Z}}

\def\E{\mathsf{E}}

\def\ii{{\rm i}}

\def\1{\tensor{1}}
\def\2{\tensor{2}}
\def\3{\tensor{3}}
\def\4{\tensor{4}}

\numberwithin{equation}{section}

\begin{document}

\title[A unifying 2d action for integrable {\large $\sigma$}-models from 4d Chern-Simons]
{A unifying 2d action for integrable\\[1mm]
{\Large $\sigma$}-models from 4d Chern-Simons theory}

\author{Francois Delduc}
\address{Univ Lyon, Ens de Lyon, Univ Claude Bernard, CNRS, Laboratoire de Physique, F-69342 Lyon, France} \email{francois.delduc@ens-lyon.fr}
\author{Sylvain Lacroix}
\address{II. Institut f\"ur Theoretische Physik, Universit\"at Hamburg, Luruper Chaussee 149, 22761 Hamburg, Germany\vspace{-2.5mm}}
\address{Zentrum f\"ur Mathematische Physik, Universit\"at Hamburg, Bundesstrasse 55, 20146 Hamburg, Germany}\email{sylvain.lacroix@desy.de}
\author{Marc Magro}
\address{Univ Lyon, Ens de Lyon, Univ Claude Bernard, CNRS, Laboratoire de Physique, F-69342 Lyon, France} \email{marc.magro@ens-lyon.fr}
\author{Beno\^{\i}t Vicedo}
\address{Department of Mathematics, University of York, York YO10 5DD, U.K.} \email{benoit.vicedo@gmail.com}

\begin{abstract}
In the approach recently proposed by K. Costello and M. Yamazaki, which is based on a four-dimensional variant of Chern-Simons theory, we derive a simple and unifying two-dimensional form for the action of many integrable $\sigma$-models which are known to admit descriptions as affine Gaudin models. This includes both the Yang-Baxter deformation and the $\lambda$-deformation of the principal chiral model. We also give an interpretation of Poisson-Lie $T$-duality in this setting and derive the action of the $\E$-model.
\end{abstract}

\maketitle

\input{epsf}

\section{Introduction}

Determining whether a given two-dimensional classical field theory is integrable is somewhat of an art.
It requires finding a connection $d + \mathcal L$ on the two-dimensional space-time $\Sigma$, valued in some complex Lie algebra $\g^\CC$, such that:
\begin{itemize}
  \item[$(a)$] It depends meromorphically on an auxiliary Riemann surface $C$,
  \item[$(b)$] It is on-shell flat,
  \item[$(c)$] The integrals of motion constructed from it are in involution.
\end{itemize}
In this article we shall restrict attention to the case when $C = \CP$. We fix a global holomorphic coordinate $z$ on $\CC \subset \CP$, called the spectral parameter.

Given the difficulty of the above task, one can turn the tables around by seeking instead to construct connections with all the above properties and, only a posteriori, identify which classical integrable field theories they correspond to.

Very recently, two different approaches for constructing integrable field theories in this way have been developed. 

\medskip

The first, proposed in \cite{Vicedo:2017cge} and further developed more recently in \cite{Delduc:2019bcl} and \cite{Lacroix:2019xeh}, is rooted in the representation theory of untwisted affine Kac-Moody algebras, or more precisely in the theory of Gaudin models associated with such algebras. The basic idea for constructing connections $d + \mathcal L$ with all of the above desired properties is, roughly, to choose a representation of a certain infinite-dimensional Lie algebra associated with the datum of the Gaudin model and apply it to the corresponding canonical element $I_A \otimes I^A$, where $\{ I^A \}$ is a basis of this Lie algebra and $\{ I_A \}$ is a basis of its dual. Specifically, under this representation we obtain \cite{Vicedo:2017cge}
\begin{equation*}
I_A \otimes I^A \; \longmapsto \; \omega (\partial_\sigma + \mathcal L_\sigma)
\end{equation*}
where $\mathcal L_\sigma$ is the component of the 1-form $\mathcal L = \mathcal L_\sigma d\sigma + \mathcal L_\tau d\tau$ along the spatial direction which we assume here to be a circle $S^1$. By construction it depends meromorphically on the spectral parameter $z$. The prefactor $\omega$ is a meromorphic 1-form which in terms of the spectral parameter $z$ is given by 
\begin{equation} \label{twist function}
\omega = \varphi(z) dz
\end{equation}
where $\varphi$ is known as the \emph{twist function}. The latter controls the form of the Poisson bracket of $\mathcal L_\sigma$ with itself which guarantees property $(c)$. Note that this approach is intrinsically formulated within the Hamiltonian framework. In particular, the temporal component $\mathcal L_\tau$ of the on-shell flat connection $d + \mathcal L$, which satisfies also $(a)$ and $(b)$ above, is induced by evolution with the Hamiltonian.

\medskip

The second approach, proposed recently in \cite{Costello:2019tri}, is based on a four-dimensional variant of Chern-Simons theory which was used in the earlier works \cite{Costello:2013zra, Costello:2013sla, Witten:2016spx, Costello:2017dso, Costello:2018gyb, Bittleston:2019gkq} to describe integrable lattice models. In fact, two types of integrable field theories were considered in \cite{Costello:2019tri}, associated with so-called order and disorder defects, respectively. We shall restrict attention to the latter class here. The action of the four-dimensional theory reads (note that we take the same overall factor as used in \cite{Vicedo:2019dej})
\begin{equation} \label{holo CS action}
S[A] = \frac{\ii}{4 \pi} \int_{\Sigma \times \CP} \omega \wedge CS(A),
\end{equation}
where $CS(A)$ is the Chern-Simons 3-form and $\omega$ is a meromorphic 1-form on $\CP$ with zeroes. The four-dimensional $\g^\CC$-valued 1-form $A = A_\sigma d\sigma + A_\tau d\tau + A_{\bar z} d\bar z$ has no $dz$-component since it drops out from the action and is therefore ignored.
To relate $A$ to a connection on $\Sigma$ one can write
\begin{equation*}
d + A = \widehat{g} (d + \mathcal L) \widehat{g}^{-1}
\end{equation*}
for some smooth $G^\CC$-valued function $\widehat{g}$ on $\Sigma \times \CP$ and 1-form $\mathcal L = \mathcal L_\sigma d\sigma + \mathcal L_\tau d\tau$. The equations of motion derived from the action \eqref{holo CS action} ensure that $\mathcal L$ satisfies both properties $(a)$ and $(b)$. Crucially, these are accompanied by boundary equations of motion for the values of $A_\sigma$ and $A_\tau$ at the poles of $\omega$. What determines the integrable field theory in this approach is then the choice of boundary conditions imposed on $A_\sigma$ and $A_\tau$ to ensure these boundary equations of motion hold.

It was shown recently in \cite{Vicedo:2019dej} that the two approaches outlined above are closely related. In particular, the Poisson bracket of $\mathcal L_\sigma$ with itself derived from a canonical analysis of the action \eqref{holo CS action} coincides with the non-ultralocal Poisson algebra obtained in the affine Gaudin model approach, where the 1-forms $\omega$ in both approaches are identified.
It follows that the connection $d + \mathcal L$ constructed using the Chern-Simons approach of \cite{Costello:2019tri} also satisfies property $(c)$, as required.

\medskip

The purpose of this article is to show that many of the integrable $\sigma$-models which had previously been described as realisations of affine Gaudin models can equally be described using the action \eqref{holo CS action}. Specifically, we identify the boundary conditions on the 1-form $A$ which give rise to: the principal chiral model with WZ-term (already covered in \cite{Costello:2019tri}), the homogeneous Yang-Baxter deformation of the principal chiral model, the Yang-Baxter $\sigma$-model with WZ-term, the $\lambda$-deformation of the principal chiral model and the bi-Yang-Baxter $\sigma$-model.

More precisely, we will suppose that the 1-form $\omega$ has at most double poles and consider three general classes of boundary conditions that can be imposed on the 1-form $A$ at the set $\bm z$ of poles of $\omega$. These are determined by a choice of Lagrangian subalgebra of either the semi-direct product $\g \ltimes \g_{\rm ab}$, where $\g$ is a real form of $\g^\CC$ and $\g_{\rm ab}$ is an abelian copy of $\g$, the direct sum $\g \oplus \g$ or the complexification $\g^\CC$. They are respectively imposed at a real double pole, at a pair of real simple poles or at a pair of complex conjugate simple poles.

One of the main results of the present paper, Theorem \ref{thm: 2d action}, is that if we impose any combination of the above three types of boundary conditions on $A$, then the four-dimensional action \eqref{holo CS action} reduces to the two-dimensional action
\begin{equation} \label{2d action intro}
S[ \{ g_x\}_{x \in \bm z} ] = \frac{1}{2} \sum_{x \in \bm z} \int_{\Sigma} \langle \res_x \omega \wedge \mathcal L, g_x^{-1} d g_x \rangle - \frac{1}{2} \sum_{x \in \bm z} (\res_x \omega) I_{\rm WZ}[g_x],
\end{equation}
where the two-dimensional field $g_x : \Sigma \to G$ is defined as the restriction $\widehat{g}|_{\Sigma \times \{ x \}}$ for all $x \in \bm z$ and $I_{\rm WZ}[g_x]$ denotes the corresponding Wess-Zumino term.

The meromorphic 1-form $\mathcal L$ can be expressed in terms of the set of fields $\{ g_x \}_{x \in \bm z}$ by solving the boundary conditions on $A$, so that the action is then a functional of these fields only. By construction, the equations of motion for these fields obtained by varying \eqref{2d action intro} is equivalent to the flatness of the connection $d + \mathcal L$.

The two-dimensional action \eqref{2d action intro} unifies the actions of many integrable $\sigma$-models which had previously been described in the affine Gaudin model approach.

\medskip

We also give an interpretation of Poisson-Lie $T$-duality in this context as arising in the case when the Lagrangian subalgebra of either $\g \oplus \g$ or $\g^\CC$ belongs to a Manin triple, \emph{i.e.} there is a complementary Lagrangian subalgebra in $\g \oplus \g$ or $\g^\CC$.

Finally, we also consider a fourth kind of natural boundary condition on $A$ imposed at a pair of simple poles, and for which the two-dimensional action \eqref{2d action intro} also holds. Imposing this boundary condition, we recover the action for the $\E$-model also from \eqref{2d action intro}. We stress, however, that this particular example is on a different footing to all of the others considered in this paper since the 1-form $\mathcal L$ in this case vanishes on-shell and so trivially satisfies condition $(b)$ above.

\subsubsection*{Acknowledgements}

This work is partially supported by the French Agence Nationale de la Recherche (ANR) under grant
 ANR-15-CE31-0006 DefIS. The work of S.L. is funded by the Deutsche Forschungsgemeinschaft (DFG, German Research Foundation) under Germany's Excellence Strategy - EXC 2121 ``Quantum Universe'' - 390833306.

\section{The four-dimensional action}

Let $G^\CC$ be a complex semisimple Lie group with Lie algebra $\g^\CC$, on which we fix a choice of non-degenerate invariant symmetric bilinear form $\langle \cdot, \cdot \rangle : \g^\CC \times \g^\CC \to \CC$.

Let $\CP \coloneqq \CC \cup \{ \infty \}$ denote the Riemann sphere. We shall fix a choice of global holomorphic coordinate $z$ on $\CC \subset \CP$.

\subsection{Bulk and boundary equations of motion} \label{sec: CS action}

Consider the action \eqref{holo CS action} where $\omega$ is a meromorphic $1$-form on $\CP$ and the Chern-Simons 3-form for the 1-form $A = A_\sigma d\sigma + A_\tau d\tau + A_{\bar z} d\bar z$ is given by
\begin{equation*}
CS(A) = \langle A, d A + \mbox{\small $\frac{2}{3}$} A \wedge A \rangle = \big\langle A, d A + \mbox{\small $\frac{1}{3}$} [A \wedge A] \big\rangle.
\end{equation*}
The second equality uses the fact that $B \wedge B = \ha [B \wedge B]$ for any $\g^\CC$-valued $1$-form $B$. Note also that for any $\g^\CC$-valued $1$-forms $B, C$ and $D$ we have
\begin{equation} \label{ad inv bil}
\langle B, [C \wedge D] \rangle = \langle C, [D \wedge B] \rangle
\end{equation}
by the invariance and symmetry of the bilinear form $\langle \cdot, \cdot \rangle$.

\medskip

Varying the action \eqref{holo CS action} with respect to the field $A$ we find
\begin{equation*}
\delta S[A] = \frac{\ii}{2 \pi} \int_{\Sigma \times \CP} \omega \wedge \langle \delta A, F(A) \rangle - \frac{\ii}{4 \pi} \int_{\Sigma \times \CP} d\omega \wedge \langle A, \delta A \rangle,
\end{equation*}
where $F(A) \coloneqq dA + A \wedge A$. The last term comes from applying Stokes's theorem and we have removed the boundary term using the fact that $A$ vanishes at the boundary of $\Sigma \times \CP$. The variation of the action therefore vanishes provided that
\begin{subequations} \label{eom and bc}
\begin{align}
\label{eom} \omega \wedge F(A) &= 0, \\
\label{bc} d\omega \wedge \langle A, \delta A \rangle &= 0.
\end{align}
\end{subequations}
The first equation \eqref{eom} is the bulk equation of motion, while the second equation \eqref{bc} we will refer to as the boundary equation of motion since $d\omega$ is a distribution supported at the set $\bm z$ of poles of $\omega$ (see the proof of Lemma \ref{lem: bc explicit} below).

More explicitly, the $\bar z$-, $\tau$- and $\sigma$-components of the bulk equation \eqref{eom} read
\begin{subequations} \label{eoms}
\begin{align}
\label{eom1} \partial_\sigma A_\tau - \partial_\tau A_\sigma + [A_\sigma, A_\tau] &= 0, \\
\label{eom2} \omega \big( \partial_{\bar z} A_\sigma - \partial_\sigma A_{\bar z} + [A_{\bar z}, A_\sigma] \big) &= 0, \\
\label{eom3} \omega \big( \partial_{\bar z} A_\tau - \partial_\tau A_{\bar z} + [A_{\bar z}, A_\tau] \big) &= 0.
\end{align}
\end{subequations}
We have kept the factor of $\omega$ in the last two equations since $\partial_{\bar z} A_\sigma$ and $\partial_{\bar z} A_\tau$ may be distributions on $\CP$, with support at the zeroes of $\omega$.

In order to rewrite the boundary equation of motion \eqref{bc} more explicitly, we begin by introducing some notation. Let $\xi_x$ be a local holomorphic coordinate around $x \in \bm z$. Explicitly $\xi_x = z - x$ for $x \in \bm z \setminus \{ \infty \}$ and $\xi_\infty = z^{-1}$ for the point at infinity.
It will also be convenient to introduce the shorthand notation $f|_x \coloneqq f|_{\Sigma \times \{ x \}}$ for the function on $\Sigma$ obtained by evaluating any function $f$ on $\Sigma \times \CP$ at $x \in \CP$.

\begin{lemma} \label{lem: bc explicit}
The boundary equation of motion \eqref{bc} can be rewritten as
\begin{equation} \label{bc explicit general}
\sum_{x \in \bm z} \sum_{p \geq 0} (\res_x \xi_x^p \omega) \epsilon_{ij} \frac{1}{p!} \partial^p_{\xi_x} \langle A_i, \delta A_j \rangle \big|_x = 0,
\end{equation}
where there is an implicit sum over the repeated space-time indices $i, j = \tau, \sigma$.
\begin{proof}
The pole part of the 1-form $\omega$ at each $x \in \bm z$ can be expressed as
\begin{equation} \label{omega pole part}
\sum_{p \geq 0} \frac{k^{(x)}_p}{\xi_x^{p+1}} d\xi_x
\end{equation}
in the local variable $\xi_x$ at $x$, where $k^{(x)}_p \coloneqq \res_x \xi_x^p \omega$. Note that this also deals with the point at infinity if $\infty \in \bm z$. Concretely, since $\xi_\infty = z^{-1}$ is the local variable at infinity this means that the pole part of $\omega$ there takes the form $- \sum_{p \geq 0} k^{(\infty)}_p z^{p-1} dz$. We then have
\begin{equation*}
d \omega = 2 \pi \ii \sum_{x \in \bm z} \sum_{p \geq 0} \frac{k^{(x)}_p}{p!} (-1)^{p} \partial^p_{\xi_x} \delta_{\xi_x 0} d\xi_x \wedge d\bar \xi_x,
\end{equation*}
where $\delta_{\xi_y 0}$ denotes the Dirac $\delta$-distribution at $y$, with the property that
\begin{equation*}
\int_{\CP} d\xi_y \wedge d\bar \xi_y \delta_{\xi_y 0} f = f|_y
\end{equation*}
for any smooth function $f : \CP \to \CC$.

Integrating $d\omega \wedge \langle A, \delta A \rangle = \epsilon_{ij} \langle A_i, \delta A_j \rangle d\omega \wedge d\sigma \wedge d\tau$ over a small open neighbourhood of $\bm z \subset \CP$ using the above expression for $d\omega$ gives the desired result.
\end{proof}
\end{lemma}

When the 1-form $\omega$ has at most double poles, which is the case we shall focus on in the present paper, the boundary equation of motion \eqref{bc explicit general} simply reads
\begin{equation} \label{bc explicit}
\sum_{x \in \bm z} (\res_x \omega) \epsilon_{ij} \langle A_i|_x, \delta A_j|_x \rangle + \sum_{x \in \bm z} (\res_x \xi_x \omega) \epsilon_{ij} \partial_{\xi_x} \langle A_i, \delta A_j \rangle\big|_x = 0.
\end{equation}

Following the approach of~\cite{Costello:2017dso, Costello:2019tri}, we will impose appropriate boundary conditions on the 1-form $A$ to ensure that the boundary equation of motion \eqref{bc explicit} is satisfied. Let us note that for a given meromorphic 1-form $\omega$, different boundary conditions can be chosen, leading to different $\sigma$-models. We therefore postpone the detailed description of the various boundary conditions we shall consider until \S\ref{sec: boundary conditions}, concentrating for the time being on aspects which are common to all these choices.

\subsection{Gauge transformations}\label{sec: gauge}

The group consisting of smooth $G^\mathbb{C}$-valued functions $u$ on $\Sigma\times\CP$ acts on the space of $\g^\CC$-valued connections $d+A$, considered in \S\ref{sec: CS action}, by formal gauge transformations
\begin{equation}\label{formal gauge}
d+A \longmapsto d+A^u \coloneqq u(d+A)u^{-1} = d - duu^{-1} + uAu^{-1} .
\end{equation}
Such transformations act on the curvature of $A$ by conjugation, namely
\begin{equation}
F(A^u) = u F(A) u^{-1}.
\end{equation}
Thus, they are symmetries of the bulk equation of motion \eqref{eom}. However, they are in general not symmetries of the boundary equation of motion \eqref{bc}. In the rest of this article, we will use the term ``gauge transformation'' to refer to the transformations $A \mapsto A^u$ which preserve the boundary conditions imposed on the field $A$ at the poles $\bm z$ of $\omega$, while keeping the denomination of ``formal gauge transformation'' to describe the most general ones. In particular, only gauge transformations leave the action \eqref{holo CS action} invariant and can thus be interpreted as local symmetries of the model.

\subsection{Lax connection}

In order to connect $A$ with the Lax connection of an integrable $\sigma$-model one should work in a formal gauge where the $d\bar z$-component vanishes.

Indeed, if we denote by $\mathcal L$ the 1-form $A$ in such a formal gauge then $\mathcal L$ would only have components along $d\sigma$ and $d\tau$, it would be on-shell flat by the first bulk equation of motion \eqref{eom1} and its dependence on $\CP$ would be meromorphic by virtue of the remaining two bulk equations of motion \eqref{eom2} and \eqref{eom3}. These are exactly the properties of a Lax connection of an integrable $\sigma$-model.

It is important to note that $\mathcal L$ is related to $A$ only by a \emph{formal} gauge transformation \eqref{formal gauge} which need not preserve the boundary conditions imposed on $A$. In particular, one cannot compute the value of the action \eqref{holo CS action} in this formal gauge. However, recall from \S\ref{sec: gauge} that, crucially, formal gauge transformations preserve the bulk equations \eqref{eoms}: this is what allowed us to interpret $\mathcal{L}$ as a Lax connection in the previous paragraph.

\medskip

Let us be more explicit about the construction sketched above. Finding the formal gauge mentioned in the previous paragraph means writing the 1-form $A$ in the form
\begin{equation} \label{A holo gauge}
A = - d \widehat{g} \widehat{g}^{-1} + \widehat{g} \mathcal L \widehat{g}^{-1},
\end{equation}
for some smooth function $\widehat{g} : \Sigma \times \CP \to G^\CC$, denoted by $\widehat{\sigma}^{-1}$ in \cite{Costello:2019tri}, and where $\mathcal L \coloneqq \mathcal L_\sigma d\sigma + \mathcal L_\tau d\tau$ has no $d\bar z$-component, \emph{i.e.} $\mathcal L_{\bar z} = 0$.

Substituting \eqref{A holo gauge} into the bulk equation of motion \eqref{eom1} implies that $\mathcal L$ is on-shell flat, while substituting it into \eqref{eom2} and \eqref{eom3} tells us that
\begin{equation} \label{L mero}
\omega \wedge \partial_{\bar z} \mathcal L  = 0.
\end{equation}
It follows from \eqref{L mero} that $\mathcal L$ is meromorphic with poles at the zeroes of $\omega$, with the order of each pole of $\mathcal L$ being at most equal to the multiplicity of the corresponding zero of $\omega$. In other words, $\omega \wedge \mathcal L$ has the same poles as $\omega$ and of the same order.

It is important to note here that there is, in fact, a large freedom in choosing a smooth $G^\CC$-valued field $\widehat{g}$ with the property \eqref{A holo gauge}. Since this will be a crucial point for us, we postpone its detailed discussion until \S\ref{sec: freedom} below.

\medskip

We can be more explicit about the pole structure of $\mathcal L$ following \cite{Costello:2019tri}, by making the choice that the singularity at each zero of $\omega$ lies only in one component of $\mathcal L$ so that $\omega \wedge CS(A)$ is regular.
Let $\bm \zeta$ denote the set of zeroes of the 1-form $\omega$ which we assume to be simple. We will allow the meromorphic 1-form $\mathcal L$ to have the form
\begin{equation} \label{L pole structure}
\mathcal L = \sum_{y \in \bm \zeta} V^y \xi_y^{-1} d\sigma_y + U_\sigma d\sigma + U_\tau d\tau
\end{equation}
where $\xi_y$ is the local coordinate at $y$ and $U_\tau, U_\sigma, V^y : \Sigma \to \g^\CC$ are smooth functions for each $y \in \bm \zeta$. Here, each $\sigma_y$ for $y \in \bm \zeta$ is a linear combination of $\sigma$ and $\tau$.

The situation considered in \cite{Costello:2019tri} corresponds to the case where $\sigma_y = w = \ha ( \tau + \ii \sigma)$ for some of the zeroes $y \in \bm \zeta$ and $\sigma_y = \bar w = \ha(\tau - \ii \sigma)$ for the others. This is the natural choice to obtain Euclidian invariant theories (see Remark \ref{rem: relativistic} below). Since we are interested in Lorentz invariant theories rather than in Euclidean invariant ones, we will instead always (with the exception of the discussion in \S\ref{sec: E-model}) make the choice $\sigma_y = \sigma^+$ for some of the zeroes $y \in \bm z$ and $\sigma_y = \sigma^-$ for the other zeroes, where $\sigma^\pm \coloneqq \ha (\tau \pm \sigma)$ are the light-cone coordinates.

\begin{remark} \label{rem: relativistic}
The form \eqref{L pole structure} is consistent with the derivation of integrable $\sigma$-model actions from their descriptions as affine Gaudin models \cite{Delduc:2019bcl}. Indeed, the spatial and temporal components of the Lax connection of an affine Gaudin model are given by very similar expressions (see \cite[(2.39) \& (2.40)]{Delduc:2019bcl} and \cite[Theorem 2.1]{Lacroix:2019xeh} for details), namely we should have
\begin{equation*}
\mathcal L = \bigg( \sum_{y \in \bm \zeta} V^y \xi_y^{-1} + U_\sigma \bigg) d\sigma + \bigg( \sum_{y \in \bm \zeta}\epsilon_y V^y \xi_y^{-1} + U_\tau \bigg) d\tau,
\end{equation*}
for some fixed $\epsilon_y \in \CC$ for all $y \in \bm \zeta$. This is equivalent to \eqref{L pole structure} with $d\sigma_y = d\sigma + \epsilon_y d\tau$.

It was also shown in \cite{Delduc:2019bcl, Lacroix:2019xeh} that for the affine Gaudin model to describe a \emph{relativistic} integrable $\sigma$-model we should take $\epsilon_y = \pm 1$ for each $y \in \bm \zeta$. This analysis can be generalised to show that the theory is Euclidean invariant if $\epsilon_y = \pm \ii$ for each $y \in \bm \zeta$. This was precisely the choice made in \cite{Costello:2019tri}, \emph{i.e.} $\sigma_y = w, \bar w$.
\end{remark}

\subsection{Action}

We will now express the action \eqref{holo CS action} in terms of $\widehat{g}$ and $\mathcal L$.

\begin{lemma}
Under a formal gauge transformation as in \eqref{A holo gauge}, the Chern-Simons 3-form transforms as
\begin{equation} \label{CS A simplified}
CS(A) = \langle \mathcal L, d \mathcal L \rangle + d \langle \widehat{g}^{-1} d \widehat{g}, \mathcal L \rangle + \mbox{\small $\frac{1}{3}$} \langle \widehat{g}^{-1} d \widehat{g}, \widehat{g}^{-1} d \widehat{g} \wedge \widehat{g}^{-1} d \widehat{g} \rangle.
\end{equation}
\begin{proof}
The behaviour of the Chern-Simons 3-form under formal gauge transformations,
\begin{equation*}
CS(A) = CS(\mathcal L) + d \langle \widehat{g}^{-1} d \widehat{g}, \mathcal L \rangle + \mbox{\small $\frac{1}{3}$} \langle \widehat{g}^{-1} d \widehat{g}, \widehat{g}^{-1} d \widehat{g} \wedge \widehat{g}^{-1} d \widehat{g} \rangle,
\end{equation*}
is well known. In the present context, since the 1-form $\mathcal L$ only has components along $d\sigma$ and $d\tau$, we have $CS(\mathcal L) = \langle \mathcal L, d \mathcal L \rangle$ from which we deduce \eqref{CS A simplified}. Since this is an essential result on which the derivation of the two-dimensional action in \S\ref{sec: 2d models} rests, we recall its proof in detail below for completeness.

\medskip

Following \cite{Costello:2019tri}, it is convenient to use the shorthand notation $A = \widehat{A} + A'$ with
\begin{equation*}
\widehat{A} \coloneqq - d \widehat{g} \widehat{g}^{-1}, \qquad
A' \coloneqq \widehat{g} \mathcal L \widehat{g}^{-1}.
\end{equation*}
We have the identity (valid for \emph{any} $1$-form $A$ decomposed as a sum $A = \widehat{A} + A'$)
\begin{align} \label{CS decompose}
CS(A) &= \langle \widehat{A} + A', d\widehat{A} + dA' \rangle + \mbox{\small $\frac{1}{3}$} \langle \widehat{A} + A', [\widehat{A} \wedge \widehat{A}] + 2 [A' \wedge \widehat{A}] + [A' \wedge A'] \rangle \notag \\
&= CS(\widehat{A}) + 2 \langle A', F(\widehat{A}) \rangle - d \langle \widehat{A}, A' \rangle + \langle \widehat{A}, [A' \wedge A'] \rangle + CS(A'),
\end{align}
where in the second line we have made use of \eqref{ad inv bil} to rearrange terms. This is to be compared with \cite[(8.8)]{Costello:2019tri}, noting that $[A' \wedge A'] = 2 A' \wedge A'$.

The second term on the right hand side of \eqref{CS decompose} vanishes by virtue of the fact that $F(\widehat{A}) = 0$, since $\widehat{A}$ is formally pure gauge.
On the other hand
\begin{align*}
CS(A') &= \langle A', d A' \rangle = \big\langle \widehat{g} \mathcal L \widehat{g}^{-1}, \big[ d\widehat{g} \widehat{g}^{-1} \wedge \widehat{g} \mathcal L \widehat{g}^{-1} \big] \big\rangle + \langle \mathcal L, d \mathcal L \rangle\\
&= - \langle A', [\widehat{A} \wedge A'] \rangle + \langle \mathcal L, d \mathcal L \rangle = - \langle \widehat{A}, [A' \wedge A'] \rangle + \langle \mathcal L, d \mathcal L \rangle.
\end{align*}
where in the first equality we have used $\langle A', A' \wedge A' \rangle = 0$ which follows using the fact that $A'$ only has $d\sigma$- and $d\tau$-components. The second and third equalities are by definition of $\widehat{A}$ and $A'$ while the last equality follows from \eqref{ad inv bil}. Finally, we have
\begin{align*}
CS(\widehat{A}) &= \langle \widehat{A}, d \widehat{A} \rangle + \mbox{\small $\frac{2}{3}$} \langle \widehat{A}, \widehat{A} \wedge \widehat{A} \rangle = - \mbox{\small $\frac{1}{3}$} \langle \widehat{A}, \widehat{A} \wedge \widehat{A} \rangle.
\end{align*}
The second equality here uses the fact that $F(\widehat{A}) = 0$ so that $d\widehat{A} = - \widehat{A} \wedge \widehat{A}$.
Putting all of the above together we obtain the desired identity \eqref{CS A simplified}.
\end{proof}
\end{lemma}

\begin{lemma} \label{lem: CSL zero}
For $\mathcal L$ of the form \eqref{L pole structure} we have $\omega \wedge \langle \mathcal L, d \mathcal L \rangle = 0$.
\begin{proof}
It follows from the explicit form \eqref{L pole structure} of the Lax connection that
\begin{equation*}
\omega \wedge \langle \mathcal L, d \mathcal L \rangle =  -2 \pi \ii \sum_{y \in \bm \zeta} \omega \wedge \big\langle \mathcal L, V^y \delta_{\xi_y 0} d\bar \xi_y \wedge d\sigma_y \big\rangle.
\end{equation*}

Consider each term in the sum over $y \in \bm \zeta$ individually. Since this already contains an explicit $d\sigma_y$, the corresponding term in $\mathcal L$ which is singular at $y$ cannot contribute. Thus only the terms which are regular at $y$ can contribute from $\mathcal L$. On the other hand, since $y$ is a simple zero of $\omega$ it follows that $\omega \, \delta_{\xi_y 0} = 0$. Thus each term in the above sum over $y \in \bm \zeta$ vanishes, as required.
\end{proof}
\end{lemma}

Substituting \eqref{CS A simplified} into the action \eqref{holo CS action} and using Lemma \ref{lem: CSL zero} we thus obtain
\begin{align} \label{action hat g}
S[A] &=  \frac{\ii}{12 \pi} \int_{\Sigma \times \CP} \omega \wedge \langle \widehat{g}^{-1} d \widehat{g}, \widehat{g}^{-1} d \widehat{g} \wedge \widehat{g}^{-1} d \widehat{g} \rangle \notag\\
&\qquad\qquad\qquad + \frac{\ii}{4 \pi} \int_{\Sigma \times \CP} d\omega \wedge \langle \widehat{g}^{-1} d \widehat{g}, \mathcal L \rangle,
\end{align}
where in the second line we have used Stokes's theorem and the fact that all fields are assumed to vanish at the boundary of $\Sigma \times \CP$ to get rid of the boundary term.

\subsection{Reality conditions}

The action \eqref{holo CS action} is a functional of the complex valued 1-forms $\omega$ and $A$. Without imposing conditions on $\omega$ and $A$ it is certainly not real.

However, we will want to use this four-dimensional theory to construct the actions of two-dimensional integrable $\sigma$-models. In order to ensure that the latter are all real, we will impose suitable reality conditions on the 1-forms $\omega$ and $A$ so as to make the four-dimensional action \eqref{holo CS action} real itself.

\medskip

Let $\tau : \g^\CC \to \g^\CC$ be an anti-linear involutive automorphism of the complex Lie algebra $\g^\CC$. It provides $\g^\CC$ with an action of the cyclic group $\ZZ_2$. Its fixed point subset is a real Lie subalgebra $\g$ of $\g^\CC$, regarded itself as a real Lie algebra. The anti-linear involution $\tau$ is compatible with the bilinear form on $\g^\CC$ in the sense that
\begin{equation} \label{tau bilinear}
\overline{\langle B, C \rangle} = \langle \tau B, \tau C \rangle
\end{equation}
for any $B, C \in \g^\CC$. We will also denote by $\tau$ its lift to an involutive automorphism $\tau : G^\CC \to G^\CC$ of the Lie group $G^\CC$ and denote by $G$ its fixed point real subgroup.

Complex conjugation $z \mapsto \bar z$ on $\CC \subset \CP$ defines an involution $\mu_{\rm t} : \CP \to \CP$, which also provides $\CP$ with a $\ZZ_2$-action. We will require both the 1-forms $\omega$ and $A$ to be equivariant under this action of $\ZZ_2$ in the sense that
\begin{equation} \label{equi omega A}
\overline{\omega} = \mu_{\rm t}^\ast \omega, \qquad
\tau A = \mu_{\rm t}^\ast A.
\end{equation}
Concretely, in terms of the twist function $\varphi$, defined from $\omega$ in \eqref{twist function}, the first condition simply states that $\overline{\varphi(z)} = \varphi(\bar z)$.

\begin{lemma}
The reality conditions \eqref{equi omega A} ensure that the action \eqref{holo CS action} is real.
\begin{proof}
We have
\begin{align} \label{real action}
\overline{S[A]} &= -\frac{\ii}{4 \pi} \int_{\Sigma \times \CP} \overline{\omega} \wedge CS(\tau A)
= -\frac{\ii}{4 \pi} \int_{\Sigma \times \CP} \mu_{\rm t}^\ast \omega \wedge CS(\mu_{\rm t}^\ast A) \notag\\
&= -\frac{\ii}{4 \pi} \int_{\Sigma \times \CP} \mu_{\rm t}^\ast (\omega \wedge CS(A))
= -\frac{\ii}{4 \pi} \int_{\Sigma \times \mu_{\rm t} \CP} \omega \wedge CS(A) = S[A],
\end{align}
where in the first equality we used the fact that
\begin{equation*}
\overline{CS(A)} = \overline{\big\langle A, d A + \mbox{\small $\frac{1}{3}$} [A \wedge A] \big\rangle}
= \big\langle \tau A, d (\tau A) + \mbox{\small $\frac{1}{3}$} [\tau A \wedge \tau A] \big\rangle = CS(\tau A).
\end{equation*}
In the middle step here we have used both the identity \eqref{tau bilinear} and the fact that $\tau$ is an automorphism of $\g^\CC$. The second step in \eqref{real action} is by the equivariance property \eqref{equi omega A} of $\omega$ and $A$. The very last step in \eqref{real action} uses the fact that $\mu_{\rm t}$ has the effect of conjugating the complex structure on $\CP$ and thus also of reversing its orientation. Concretely, the integral over $\mu_{\rm t} \CP$ with measure $d\bar z \wedge dz$ is equal to the integral over $\CP$ with measure $d z \wedge d\bar z$.
\end{proof}
\end{lemma}

Upon writing the 1-form $A$ as in \eqref{A holo gauge}, to satisfy its equivariance property \eqref{equi omega A} we will impose the equivariance property
\begin{equation} \label{equi hatg L}
\tau \widehat{g} = \mu_{\rm t}^\ast \widehat{g}, \qquad \tau \mathcal L = \mu_{\rm t}^\ast \mathcal L.
\end{equation}
for the function $\widehat{g} : \Sigma \times \CP \to G^\CC$ and $\g^\CC$-valued 1-form $\mathcal L = \mathcal L_\sigma d\sigma + \mathcal L_\tau d\tau$.

\section{Integrable $\sigma$-model actions} \label{sec: 2d action}

\subsection[Freedom in the choice of $\widehat{g}$]{Freedom in the choice of $\bm{\widehat{g}}$} \label{sec: freedom}

Notice that \eqref{A holo gauge} is equivalent to saying that $A_{\bar z}$ is of the form
\begin{equation} \label{A holo gauge 2}
A_{\bar z} = - \partial_{\bar z} \widehat{g} \widehat{g}^{-1}.
\end{equation}
The smooth function $\widehat{g} : \Sigma \times \CP \to G^\CC$ in this expression is by no means unique.

\medskip

On the one hand, we can multiply it on the right by an arbitrary smooth function $h : \Sigma \to G$ since we have
\begin{equation} \label{ghat change 1}
A_{\bar z} = - \partial_{\bar z} (\widehat{g} h) (\widehat{g} h)^{-1},
\end{equation}
which is still of the form \eqref{A holo gauge 2}. In order to preserve the equivariance of $\widehat{g}$ in \eqref{equi hatg L} we need $h$ to take values in the real subgroup $G \subset G^\CC$ so that $\tau h = h$.

Note that such a transformation $\widehat{g}\mapsto\widehat{g}h$ does not modify $A_{\bar z}$ and is thus a redundancy in the definition \eqref{A holo gauge 2} of $\widehat{g}$ in terms of $A_{\bar z}$. Recall also that this definition was obtained as the $d\bar z$-component of \eqref{A holo gauge} and that the corresponding $d\tau$- and $d\sigma$-components serve as a definition of the Lax connection $\mathcal{L}$ in terms of $A_\tau$, $A_\sigma$ and $\widehat{g}$. One easily checks that for fixed $A$, the redundancy $\widehat{g}\mapsto\widehat{g}h$ in the definition of $\widehat{g}$ corresponds to the transformation
\begin{equation}\label{2dgauge}
\mathcal{L} \longmapsto h^{-1} dh + h^{-1} \mathcal{L}h
\end{equation}
on $\mathcal{L}$.
This is a two-dimensional gauge transformation of the Lax connection $\mathcal{L}$. It is well known that such a freedom on $\mathcal{L}$ is always allowed in any integrable field theory, as it preserves its on-shell flatness. 
\medskip

On the other hand, we can also perform a gauge transformation on the connection $A$ by a smooth function $u : \Sigma \times \CP \to G^\CC$ since the $d\bar z$-component of the gauge transformed connection
\begin{equation} \label{gauge transformation u}
A^u = - d u u^{-1} + u A u^{-1}
\end{equation}
is still of the form \eqref{A holo gauge 2}, explicitly
\begin{equation} \label{ghat change 2}
A^u_{\bar z} = - \partial_{\bar z}(u \widehat{g}) (u \widehat{g})^{-1}.
\end{equation}
However, it is important to note that $u$ cannot be completely arbitrary here. Indeed, the gauge transformation by $u$ must also preserve the boundary conditions imposed on $A$ (which is why, following the terminology of \S\ref{sec: gauge}, we call it a gauge transformation and not a formal gauge transformation). For $A^u$ to be real we must also require that $u$ be equivariant under the action of $\ZZ_2$.

Note that the transformation $\widehat{g}\mapsto u\widehat{g}$ is of a different nature than the transformation $\widehat{g}\mapsto\widehat{g}h$ considered in the previous paragraph. Indeed, the latter corresponds to a redundancy in the definition of $\widehat{g}$ in terms of $A$ and does not alter $A$ itself, while the transformation $\widehat{g}\mapsto u\widehat{g}$ corresponds to a gauge transformation on $A$. Moreover, the parameter $h$ considered in \eqref{ghat change 1} was a two-dimensional field on $\Sigma$, independent of $z$ and $\bar z$, while the parameter $u$ in \eqref{gauge transformation u} is a four-dimensional field on $\Sigma \times \CP$. Finally, let us note that contrary to the transformation $\widehat{g} \mapsto \widehat{g} h$, the gauge transformation $\widehat{g}\mapsto u\widehat{g}$ does not modify the Lax connection $\mathcal{L}$.

\subsection{Archipelago conditions} \label{sec: bc}

The action \eqref{action hat g} derived in the previous section holds for an arbitrary meromorphic differential $\omega$, in particular with poles of any order. It is, however, still four-dimensional as the original action \eqref{holo CS action}.

In order to reduce the action \eqref{action hat g} to a two-dimensional one, we will exploit the large freedom in the choice of $\widehat{g}$ discussed in \S\ref{sec: freedom}. Specifically, in the remainder of this section we will identify sufficient conditions on the function $\widehat{g}$ which guarantee that the action \eqref{action hat g} can be explicitly reduced to an action on $\Sigma$. In \S\ref{sec: boundary conditions} below we will then identify various boundary conditions for which such conditions on $\widehat{g}$ can be made to hold by using the freedom discussed in \S\ref{sec: freedom}.

\medskip

We will say that a smooth equivariant function $\widehat{g} : \Sigma \times \CP \to G^\CC$ is of \emph{archipelago type} if it satisfies the following three \emph{archipelago conditions}:
\begin{itemize}
  \item[$(i)$] $\widehat{g} = 1$ outside $\Sigma \times \bigsqcup_{x \in \bm z} U_x$ for some disjoint open discs $U_x$ around $x \in \bm z$, 
  \item[$(ii)$] $\widehat{g}_x \coloneqq \widehat{g}|_{\Sigma \times U_x}$ only depends on $\sigma$, $\tau$ and the radial coordinate $r_x \coloneqq |\xi_x|$,
  \item[$(iii)$] There is an open disc $V_x \subset U_x$ for every $x \in \bm z$ such that $g_x \coloneqq \widehat{g}|_{\Sigma \times V_x}$ only depends on $\sigma$ and $\tau$. By a slight abuse of notation we also denote its further restriction $\widehat{g}|_{\Sigma \times \{ x \}}$ to the point $x \in \bm z$ as $g_x$.
\end{itemize}

\begin{lemma} \label{lem: bc}
One can always ensure that the smooth $G^\CC$-valued function $\widehat{g}$ appearing in \eqref{A holo gauge 2} satisfies the archipelago condition $(i)$.
\begin{proof}
We will bring the function $\widehat{g}$ to a form which satisfies the archipelago condition $(i)$ by applying a suitable gauge transformation \eqref{gauge transformation u} for some smooth function $u$.

Given any disjoint open discs $U_x$ around each $x \in \bm z$ we can choose a smooth function $u : \Sigma \times \CP \to G^\CC$ such that $u = \widehat{g}^{-1}$ outside $\Sigma \times \bigsqcup_{x \in \bm z} U_x$ and $u = 1$ in some open neighbourhood of $\Sigma \times \bm z$. The latter condition is there to ensure that the gauge transformation by $u$ preserves the boundary conditions at $\bm z$. By construction, the new function $u \widehat{g}$ appearing in \eqref{ghat change 2} satisfies condition $(i)$.
\end{proof}
\end{lemma}

By contrast, the conditions $(ii)$ and $(iii)$ are not always satisfied. Whether or not $\widehat{g}$ can be made to satisfy them depends on the type of boundary conditions that are imposed on the Chern-Simons field $A$ at the poles of $\omega$ in order to satisfy \eqref{bc explicit general}.

\subsection{Two-dimensional action with WZ-terms} \label{sec: 2d models}

Suppose that $\widehat{g}$ can be chosen to be of archipelago type. We will show that the four-dimensional action \eqref{action hat g} can then be further simplified to a two-dimensional action with WZ-terms.

Consider, to begin with, the first term in the action \eqref{action hat g}. It can be written as
\begin{align*}
& \frac{\ii}{12 \pi} \int_{\Sigma \times \CP} \omega \wedge \langle \widehat{g}^{-1} d \widehat{g}, \widehat{g}^{-1} d \widehat{g} \wedge \widehat{g}^{-1} d \widehat{g} \rangle\\
&\qquad\qquad =  \frac{\ii}{12 \pi} \sum_{x \in \bm z} \int_{\Sigma \times U_x} \omega \wedge \langle \widehat{g}_x^{-1} d \widehat{g}_x, \widehat{g}_x^{-1} d \widehat{g}_x \wedge \widehat{g}_x^{-1} d \widehat{g}_x \rangle
\end{align*}
using property $(i)$ of the archipelago type function $\widehat{g}$, cf. \S\ref{sec: bc}, to localise the integral over $\CP$ to the individual discs $U_x$ around each $x \in \bm z$. 

In each disc $U_x$ centred on $x \in \bm z \setminus \{ \infty \}$ we introduce local polar coordinates $z = x + r_x e^{\ii \theta_x}$ and likewise $z = r_\infty^{-1} e^{-\ii \theta_\infty}$ in $U_\infty$ if $\infty \in \bm z$. We note that only the differential $d \theta_x$ in $dz = e^{\ii \theta_x} (dr_x + \ii r_x d \theta_x)$ contributes in the above integral for $x \in \bm z \setminus \{ \infty \}$. Indeed, since $\widehat{g}_x$ is assumed to be independent of $\theta_x$ in property $(ii)$ of the archipelago type function $\widehat{g}$, it follows that the 3-form $\langle \widehat{g}_x^{-1} d \widehat{g}_x, \widehat{g}_x^{-1} d \widehat{g}_x \wedge \widehat{g}_x^{-1} d \widehat{g}_x \rangle$ is proportional to $dr_x \wedge d\sigma \wedge d\tau$. Therefore when taking the wedge product with $\omega$, only the $d\theta_x$ component of $\omega$ can contribute. The same is true when $x = \infty$.
We can then rewrite the above integral as
\begin{align*}
&-\frac{1}{12 \pi} \sum_{x \in \bm z \setminus \{ \infty \}} \int_{\Sigma \times [0, R_x] \times [0, 2 \pi]} \!\!\!\!\!\!\! r_x e^{\ii \theta_x} \varphi\big( x + r_x e^{\ii \theta_x} \big) d\theta_x \wedge \langle \widehat{g}_x^{-1} d \widehat{g}_x, \widehat{g}_x^{-1} d \widehat{g}_x \wedge \widehat{g}_x^{-1} d \widehat{g}_x \rangle\\
&\; + \frac{1}{12 \pi} \sum_{x \in \bm z \cap \{ \infty \}} \int_{\Sigma \times [0, R_x] \times [0, 2 \pi]} \!\!\!\!\!\!\! r_x^{-1} e^{-\ii \theta_x} \varphi\big( r_x^{-1} e^{-\ii \theta_x} \big) d\theta_x \wedge \langle \widehat{g}_x^{-1} d \widehat{g}_x, \widehat{g}_x^{-1} d \widehat{g}_x \wedge \widehat{g}_x^{-1} d \widehat{g}_x \rangle,
\end{align*}
where $R_x$ is the radius of the disc $U_x$ around $x \in \bm z$. We choose an orientation on the discs $U_x$ such that $r_xdr_x\wedge d\theta_x$ is the surface element.
Performing the integrals over the angular variables $\theta_x$ for each $x \in \bm z$ we now deduce that, when $\widehat{g}$ is of archipelago type, the first term in the action \eqref{action hat g} reduces to
\begin{align*}
& \frac{\ii}{12 \pi} \int_{\Sigma \times \CP} \omega \wedge \langle \widehat{g}^{-1} d \widehat{g}, \widehat{g}^{-1} d \widehat{g} \wedge \widehat{g}^{-1} d \widehat{g} \rangle = - \frac{1}{2} \sum_{x \in \bm z} (\res_x \omega) I_{\rm WZ}[g_x].
\end{align*}
Here we introduce the standard WZ-term
\begin{equation*}
I_{\rm WZ}[g_x] \coloneqq - \frac{1}{3} \int_{\Sigma \times [0, R_x]} \langle \widehat{g}_x^{-1} d \widehat{g}_x, \widehat{g}_x^{-1} d \widehat{g}_x \wedge \widehat{g}_x^{-1} d \widehat{g}_x \rangle.
\end{equation*}
As usual, it depends only on the two-dimensional field $g_x : \Sigma \to G$ up to an additive constant which is irrelevant classically. Note that the overall minus sign in the above definition is there to match with the conventions of \cite{Delduc:2019bcl}. Indeed, the boundary of the volume $\Sigma \times [0, R_x]$ being at the origin of the interval $[0, R_x]$ accounts for this extra minus sign.

\medskip

Consider now the second term in the action \eqref{action hat g}. It can be rewritten as
\begin{equation} \label{kinetic term}
 \frac{\ii}{4 \pi} \int_{\Sigma \times \CP} d\omega \wedge \langle \widehat{g}^{-1} d \widehat{g}, \mathcal L \rangle
=  \frac{\ii}{4 \pi} \sum_{x \in \bm z} \int_{\Sigma \times V_x} d\omega \wedge \langle g_x^{-1} d g_x, \mathcal L \rangle
\end{equation}
where we have used the fact that $d\omega$ is a distribution with support $\bm z$ to localise the integral over $\CP$ to the open discs $V_x$ for each $x \in \bm z$ from property $(iii)$ of the archipelago type function $\widehat{g}$. By writing this distribution explicitly in terms of the local coordinates $\xi_x$ at each $x \in \bm z$, as in the proof of Lemma \ref{lem: bc explicit}, substituting this expression into \eqref{kinetic term} we arrive at
\begin{align*}
 \frac{\ii}{4 \pi} \int_{\Sigma \times \CP} d\omega \wedge \langle \widehat{g}^{-1} d \widehat{g}, \mathcal L \rangle
= - \frac{1}{2} \sum_{x \in \bm z} \sum_{p \geq 0} \int_{\Sigma} \frac{k^{(x)}_p}{p!} \big( \partial_{\xi_x}^p \langle g_x^{-1} d g_x, \mathcal L \rangle \big) \big|_x,
\end{align*}
where $k^{(x)}_p = \res_x \xi_x^p \omega$ for each $p \in \ZZ_{\geq 0}$ and $x \in \bm z$.

Now since $g_x$ is independent of the local coordinate $\xi_x$ on $V_x$ by property $(iii)$, it follows that $\langle g_x^{-1} d g_x, \mathcal L \rangle$ is holomorphic in a neighbourhood of the pole $x$ of $\omega$ by virtue of \eqref{L mero} and we may thus rewrite each term in the above sum over $x \in \bm z$ as a residue. Indeed, for any $\psi$ holomorphic at $x$ we have
\begin{align*}
\res_x \omega \wedge \psi = \res_x \bigg( \sum_{p \geq 0} \frac{k^{(x)}_p}{\xi_x^{p+1}} d\xi_x \wedge \sum_{q \geq 0} \frac{1}{q!} (\partial^q_{\xi_x} \psi)|_x \xi_x^q \bigg) = \sum_{p \geq 0} \frac{k^{(x)}_p}{p!} (\partial^p_{\xi_x} \psi)|_x,
\end{align*}
where in the first equality we made use of the expression \eqref{omega pole part} for the pole part of $\omega$ at $x$, as well as the Taylor expansion of $\psi$ near $x$. Finally, we thus obtain
\begin{align*}
 \frac{\ii}{4 \pi} \int_{\Sigma \times \CP} d\omega \wedge \langle \widehat{g}^{-1} d \widehat{g}, \mathcal L \rangle
&= - \frac{1}{2} \sum_{x \in \bm z} \int_{\Sigma} \res_x \big( \omega \wedge \langle g_x^{-1} d g_x, \mathcal L \rangle \big)\\
&= - \frac{1}{2} \sum_{x \in \bm z} \int_{\Sigma} \langle g_x^{-1} d g_x, \res_x \omega \wedge \mathcal L \rangle.
\end{align*}
Notice that the sign has not changed in the last line since we have moved $\omega$ past $g_x^{-1} dg_x$ but at the same time we have also reversed the orientation of the domain of integration by moving the operation $\res_x$, which is given by a contour integral over a small circle around $x$, past $g_x^{-1} dg_x$ also.

\medskip

We have thus shown the following.

\begin{theorem} \label{thm: 2d action}
If $\widehat{g}$ is of archipelago type then the action \eqref{action hat g} reduces to the sum of a two-dimensional term and a Wess-Zumino term for each point in $\bm z$, namely
\begin{equation} \label{action sigma model}
S[ \{ g_x\}_{x \in \bm z} ] = \frac{1}{2} \sum_{x \in \bm z} \int_{\Sigma} \langle \res_x \omega \wedge \mathcal L, g_x^{-1} d g_x \rangle - \frac{1}{2} \sum_{x \in \bm z} (\res_x \omega) I_{\rm WZ}[g_x],
\end{equation}
where $g_x : \Sigma \to G$ is the restriction of $\widehat{g}$ to $\Sigma \times \{ x \}$ for each $x \in \bm z$.
\end{theorem}

\begin{remark} \label{rem: L and gx}
The notation that we have used for the action in \eqref{action sigma model} suggests that it is only a functional of $\{ g_x\}_{x \in \bm z}$, even though the right hand side clearly also depends on $\mathcal L$. This is because, as we shall see in a case-by-case analysis of all the examples discussed in \S\ref{sec: examples} below, the 1-form $\mathcal L$ can always be expressed in terms of the set of fields $\{ g_x\}_{x \in \bm z}$ by solving the boundary condition imposed on $A$.
\end{remark}

It follows from the equivariance properties \eqref{equi omega A} that the set $\bm z$ of poles of $\omega$ is invariant under complex conjugation, so that $x \in \bm z$ implies $\bar x \in \bm z$. And using also \eqref{equi hatg L} we find that
\begin{equation} \label{reality res}
\overline{\res_x \omega \wedge \mathcal L} = \res_{\bar x} \omega \wedge \mathcal L, \qquad
\overline{\res_x \omega} = \res_{\bar x} \omega.
\end{equation}
Moreover, from the equivariance property \eqref{equi hatg L} it follows that for any $x \in \bm z$ we have $\tau(g_x) = g_{\bar x}$ and $\tau(\widehat{g}_x) = \widehat{g}_{\bar x}$. This, together with \eqref{reality res}, implies that the action \eqref{action sigma model} is real, as expected since it was obtained as a reduction of \eqref{holo CS action} which was real by virtue of the equivariance properties \eqref{equi omega A} imposed on $\omega$ and $A$.

\subsection{Two-dimensional gauge invariance}

Recall from the discussion in \S\ref{sec: freedom} that there is a redundancy in the definition \eqref{A holo gauge} of both the function $\widehat{g}$ and the 1-form $\mathcal L$ in terms of $A$, namely
\begin{equation*}
\widehat{g} \longmapsto \widehat{g} h, \qquad \mathcal L \longmapsto h^{-1} dh + h^{-1} \mathcal{L} h,
\end{equation*}
for an arbitary smooth function $h : \Sigma \to G$. We note that the above transformation on $\widehat{g}$ will spoil the fact that $\widehat{g}$ is of archipelago type. However, by combining it with the gauge transformation by $u$ defined in the proof of Lemma \ref{lem: bc} we are able to bring $\widehat{g} h$ back to being of archipelago type. Note that the gauge transformation by $u$ leaves invariant the 1-form $\mathcal L$ so that we obtain the combined transformation
\begin{equation} \label{gauge tr hatg L}
\widehat{g} \longmapsto u \widehat{g} h, \qquad \mathcal L \longmapsto h^{-1} dh + h^{-1} \mathcal{L} h.
\end{equation}
As $u \widehat{g} h$ is of archipelago type, the action \eqref{action sigma model} therefore holds after performing the transformation \eqref{gauge tr hatg L} and in particular it makes sense to ask whether it is invariant under such a transformation.

More precisely, in terms of the fields $\{ g_x \}_{x \in \bm z}$ appearing in the action \eqref{action sigma model}, the transformation \eqref{gauge tr hatg L} acts as
\begin{equation} \label{gauge tr gx}
g_x \longmapsto g_x h
\end{equation}
for all $x \in \bm z$. Here we used the property that $u|_x = 1$ from the proof of Lemma \ref{lem: bc}.

And as noted in Remark \ref{rem: L and gx}, in all the cases to be considered in \S\ref{sec: examples} the 1-form $\mathcal L$ will be completely fixed in terms of $\{ g_x \}_{x \in \bm z}$ by solving the boundary condition imposed on $A$. In this sense, the transformation \eqref{2dgauge} on the 1-form $\mathcal L$, \emph{i.e.} the second relation in \eqref{gauge tr hatg L}, can be seen as a consequence of \eqref{gauge tr gx}.

\begin{proposition} \label{prop: gauge-inv 2d action}
The two-dimensional action \eqref{action sigma model} is invariant under the gauge transformation \eqref{gauge tr gx} for an arbitrary smooth function $h : \Sigma \to G$.

We can fix this gauge invariance by imposing that $g_x = 1$ for some $x \in \bm z$.
\begin{proof}
We compute $S[ \{ g_x h \}_{x \in \bm z}]$ by substituting the transformations \eqref{gauge tr gx} and \eqref{2dgauge} into \eqref{action sigma model}. The first term in the action reads
\begin{align*}
&\frac{1}{2} \sum_{x \in \bm z} \int_{\Sigma} \big\langle \res_x \big( \omega \wedge (h^{-1} dh + h^{-1} \mathcal{L} h) \big), (g_x h)^{-1} d (g_x h) \big\rangle\\
&\qquad\qquad = \frac{1}{2} \sum_{x \in \bm z} \int_{\Sigma} \big\langle \res_x \big( \omega \wedge (d h h^{-1} + \mathcal L) \big), g_x^{-1} d g_x \big\rangle\\
&\qquad\qquad\qquad\qquad + \frac{1}{2} \sum_{x \in \bm z} \int_{\Sigma} \big\langle \res_x \big( \omega \wedge (h^{-1} dh + h^{-1} \mathcal{L} h) \big), h^{-1} d h \big\rangle.
\end{align*}
The second term on the right hand side vanishes because $\omega \wedge (h^{-1} dh + h^{-1} \mathcal{L} h)$ is meromorphic on $\CP$ with poles in $\bm z$, so that the sum of its residues vanishes.

On the other hand, by using the Polyakov-Wiegmann formula \cite{Polyakov-Wiegmann} we find
\begin{align*}
\frac{1}{2} \sum_{x \in \bm z} (\res_x \omega) I_{\rm WZ}[g_x h] &= 
\frac{1}{2} \sum_{x \in \bm z} (\res_x \omega) I_{\rm WZ}[g_x] + \frac{1}{2} \sum_{x \in \bm z} (\res_x \omega) I_{\rm WZ}[h]\\
&\qquad - \frac{1}{2} \sum_{x \in \bm z} (\res_x \omega) \int_\Sigma \langle g_x^{-1} dg_x, dh h^{-1} \rangle.
\end{align*}
The second term on the right hand side vanishes using the fact that $\sum_{x \in \bm z} \res_x \omega = 0$.
It now follows from combining the above that $S[ \{ g_x h \}_{x \in \bm z} ] = S[ \{ g_x \}_{x \in \bm z} ]$.
\end{proof}
\end{proposition}

\begin{remark}
In the approach to integrable $\sigma$-models based on affine Gaudin models, the gauge transformation \eqref{gauge tr gx} and its interplay with the integrable structure was studied in details in~\cite{Lacroix:2019xeh}, expanding on the description of gauge symmetries in affine Gaudin models given in~\cite{Vicedo:2017cge}. In particular, it was shown in~\cite[Proposition 2.2]{Lacroix:2019xeh} (see also \cite[(4.61)]{Vicedo:2017cge}) that the gauge transformation of the fundamental fields of the $\sigma$-model, represented here by $\{ g_x \}_{x \in \bm z}$, acts as $d+\mathcal{L} \mapsto h^{-1}(d+\mathcal{L})h$ on its Lax connection, in agreement with the situation considered in the present paper.
\end{remark}

\section{Boundary conditions} \label{sec: boundary conditions}

As already mentioned in \S\ref{sec: CS action}, we shall restrict attention in this paper to the case when $\omega$ has at most double poles, in which case the boundary conditions imposed on $A$ should ensure that \eqref{bc explicit} holds. In the list of examples discussed in \S\ref{sec: examples} we shall consider two types of boundary conditions.

The first is imposed at a double pole $x \in \bm z$ of $\omega$ and ensures that the corresponding term in the sum of \eqref{bc explicit} vanishes by itself, \emph{i.e.}
\begin{subequations}
\begin{equation} \label{bc double pole general}
(\res_x \omega) \epsilon_{ij} \langle A_i|_x, \delta A_j|_x \rangle + (\res_x \xi_x \omega) \epsilon_{ij} \partial_{\xi_x} \langle A_i, \delta A_j \rangle \big|_x = 0.
\end{equation}
For the discussion of reality conditions, we will assume for simplicity that $x$ lies on the real axis. We discuss the simplest possible boundary condition in \S\ref{sec: bc double} and then come back to more general boundary conditions that can be imposed in \S\ref{sec: bc double gen}.

The second is imposed at a \emph{pair} of simple poles $x_+, x_- \in \bm z$ of $\omega$ and ensures that the corresponding terms in the sum of \eqref{bc explicit} cancel each other out, \emph{i.e.}
\begin{equation} \label{bc simple pole general}
(\res_{x_+} \omega) \epsilon_{ij} \langle A_i|_{x_+}, \delta A_j|_{x_+} \rangle + (\res_{x_-} \omega) \epsilon_{ij} \langle A_i|_{x_-}, \delta A_j|_{x_-} \rangle = 0.
\end{equation}
\end{subequations}
There are two possibilities allowed by the reality conditions, corresponding to the case when $x_+$ and $x_-$ are both real and when they form a complex conjugate pair. These separate cases are discussed in \S\ref{sec: bc real simple} and \S\ref{sec: bc complex simple}, respectively.

In \S\ref{sec: Manin triple} we describe Poisson-Lie $T$-duality in the present context, as relating different choices of boundary conditions that can be imposed at a pair of simple poles.

\subsection{Boundary conditions at a real double pole} \label{sec: bc double}

Let $x \in \bm z$ be a real double pole of $\omega$. One way the boundary equation of motion \eqref{bc double pole general} can be satisfied is by demanding that \cite{Costello:2017dso, Costello:2019tri}
\begin{equation} \label{bc double pole}
A_i|_x = 0,
\end{equation}
for $i = \tau, \sigma$, noting that we then also have $\delta A_i|_x = 0$.

\begin{proposition} \label{prop: bc double}
Suppose that $A$ satisfies the boundary condition \eqref{bc double pole}, and we are given a field $\widehat{g}$ satisfying \eqref{A holo gauge 2} for which the archipelago condition $(i)$ holds.

Then the value of $\widehat{g}$ on the island $U_x$ can be modified, without changing its value at $x$ and its value outside $U_x$, so as to also satisfy both of the remaining two archipelago conditions $(ii)$ and $(iii)$.
\begin{proof}
This will be achieved by applying a suitable gauge transformation \eqref{gauge transformation u}
for some smooth function $u : \Sigma \times \CP \to G^\CC$, equal to $1$ on the complement of $\Sigma \times U_x$ so as not to modify the value of $\widehat{g}$ there. Note, however, that in order for $A^u$ to still satisfy the boundary condition \eqref{bc double pole} it is necessary to require that $(- \partial_i u u^{-1})|_x = 0$ for $i = \tau, \sigma$. That is, $u$ is an allowed gauge transformation parameter provided
\begin{equation} \label{allowed gt double}
\partial_i (u |_x) = 0,
\end{equation}
for $i = \tau, \sigma$.
Also, for $A^u$ to still satisfy the reality conditon \eqref{equi omega A}, we should require that $u$ be equivariant in the sense that $\tau u = \mu_{\rm t}^\ast u$.
We are thus seeking a smooth $\ZZ_2$-equivariant $G^\CC$-valued function $u$ equal to $1$ outside $\Sigma \times U_x$ and satisfying \eqref{allowed gt double}, such that $u \widehat{g}$ satisfies the archipelago conditions $(ii)$ and $(iii)$ on the island $U_x$.

Consider the smooth equivariant function $\widetilde{g} : \Sigma \times \CP \to G^\CC$ defined as follows.
Let $\widetilde{g} \coloneqq \widehat{g}$ on the complement of $\Sigma \times U_x$. Choose two open discs $D^r_x \subset D^s_x \subset U_x$ of radii $s > r > 0$ centred on $x$. Let $\widetilde{g}$ in $D^r_x$ be constant equal to $\widehat{g}|_x$, and extend it to a smooth function on $U_x$ such that $\widetilde{g} \coloneqq 1$ on the complement $U_x \setminus D^s_x$ and $\widetilde{g}$ depends only on the radial coordinate $|\xi_x|$ around $x$. More precisely, writing $\widehat{g}|_x = \exp y$ for some $y : \Sigma \to \g$, we let $\widetilde{g} \coloneqq \exp(f(|\xi_x|) y)$ where $f : [0, R_x] \to \RR$ is a smooth function equal to $1$ on $[0, r]$ and equal to $0$ on $[s, R_x]$.

By construction, $\widetilde{g}$ satisfies both of the archipelago conditions $(ii)$ and $(iii)$ on the island $U_x$. It therefore remains to show that $u = \widetilde{g} \widehat{g}^{-1} : \Sigma \times \CP \to G^\CC$ satisfies \eqref{allowed gt double} and is also $\ZZ_2$-equivariant. The latter condition is evident from the equivariance of $\widetilde{g}$ and $\widehat{g}$. On the other hand, $\partial_i (u |_x) = \partial_i \big( \widetilde{g}|_x \widehat{g}|_x^{-1} \big) = 0$ where in the second equality we used the fact that $\widetilde{g} = \widehat{g}|_x$ in $D^r_x$ and hence $\widetilde{g}|_x = \widehat{g}|_x$.
\end{proof}
\end{proposition}

\subsection{Boundary conditions at pairs of real simple poles} \label{sec: bc real simple}

Let $x_\pm \in \bm z$ be simple poles of $\omega$ with $x_\pm \in \RR$, so that in particular $\res_{x_\pm} \omega \in \RR$. Also, by the equivariance property \eqref{equi omega A} of $A$ it follows that the components $A_i|_{x_\pm}$, for $i = \tau, \sigma$, are valued in the real Lie subalgebra $\g$.

The boundary equation of motion \eqref{bc simple pole general} can then be rewritten as
\begin{equation} \label{bc opposite res}
\epsilon_{ij} \langle\!\langle (A_i|_{x_+}, A_i|_{x_-}), \delta (A_j|_{x_+}, A_j|_{x_-}) \rangle\!\rangle_{\d; x_\pm} = 0,
\end{equation}
where $\langle\!\langle \cdot, \cdot \rangle\!\rangle_{\d; x_\pm} : \d \times \d \to \RR$ denotes the non-degenerate symmetric invariant bilinear form on the Lie algebra direct sum $\d \coloneqq \g \oplus \g$, defined by
\begin{equation*}
\langle\!\langle (\ms x, \ms y), (\ms x', \ms y') \rangle\!\rangle_{\d; x_\pm} \coloneqq (\res_{x_+} \omega) \langle \ms x, \ms x' \rangle + (\res_{x_-} \omega) \langle \ms y, \ms y' \rangle
\end{equation*}
for any $\ms x, \ms y, \ms x', \ms y' \in \g$. In the special case when $\res_{x_+} \omega = - \res_{x_-} \omega$ this reduces to the usual bilinear form $\langle \ms x, \ms x' \rangle - \langle \ms y, \ms y' \rangle$ on $\d$ up to an overall factor of $\res_{x_+} \omega$.

One way of ensuring that \eqref{bc opposite res} holds is as follows. Let $(\d, \k)$ be a Manin pair, \emph{i.e.} fix a Lagrangian subalgebra $\k$ of $\d$. We recall that Lagrangian here means `maximal isotropic'. We can demand that, for $i = \tau, \sigma$,
\begin{equation} \label{bc simple pole}
(A_i|_{x_+}, A_i|_{x_-}) \in \k,
\end{equation}
noting that we will then also have $\delta (A_i|_{x_+}, A_i|_{x_-}) \in \k$. This then ensures \eqref{bc opposite res} holds by virtue of the isotropy of $\k$. The reason for using a Manin pair $(\d, \k)$ rather than just an isotropic subspace $\k$ of $\d$ will be explained shortly.

Let $K$ denote the subgroup of $D = G \times G$ with Lie algebra $\k \subset \d$.

\begin{proposition} \label{prop: bc real simple}
Suppose that $A$ satisfies the boundary condition \eqref{bc simple pole}, and we are given a field $\widehat{g}$ satisfying \eqref{A holo gauge 2} for which the archipelago condition $(i)$ holds.

Then the value of $\widehat{g}$ on the islands $U_{x_\pm}$ can be modified, without changing its value outside, so as to also satisfy the remaining archipelago conditions $(ii)$ and $(iii)$.

Furthermore, the value $(g_{x_+}, g_{x_-}) : \Sigma \to D$ of the archipelago type function $\widehat{g}$ at the pair of points $x_\pm$ can be adjusted using $(g_{x_+}, g_{x_-}) \mapsto a (g_{x_+}, g_{x_-})$ for any smooth function $a : \Sigma \to K$.
\begin{proof}
We will find a gauge transformation \eqref{gauge transformation u}
for some suitable equivariant $u : \Sigma \times \CP \to G^\CC$ equal to $1$ outside $\Sigma \times (U_{x_+} \sqcup U_{x_-})$ such that $u \widehat{g}$ also satisfies the archipelago conditions $(ii)$ and $(iii)$ on $U_{x_\pm}$.

Evaluating \eqref{gauge transformation u} at the pair of points $x_\pm$ we see that, for $i= \tau, \sigma$,
\begin{align*}
(A_i^u|_{x_+}, A_i^u|_{x_-}) &= - \big( (\partial_i u u^{-1})|_{x_+}, (\partial_i u u^{-1})|_{x_-} \big)\\
&\qquad\qquad + (u|_{x_+}, u|_{x_-}) (A_i|_{x_+}, A_i|_{x_-}) (u|_{x_+}, u|_{x_-})^{-1}.
\end{align*}
The gauge transformation is allowed provided that this still takes values in $\k$, so that $A^u$ still satisfies the boundary condition \eqref{bc simple pole}. For this, it is sufficient to ensure that both terms on the right hand side above take values in $\k$. Therefore, we will demand that our gauge transformation parameter $u$ should be such that
\begin{equation} \label{allowed u simple}
(u|_{x_+}, u|_{x_-}) \in K.
\end{equation}
Note that this then also implies $\big( (\partial_i u u^{-1})|_{x_+}, (\partial_i u u^{-1})|_{x_-} \big)  \in \k$ for $i = \tau, \sigma$. This is where we had to use the fact that $\k$ is a subalgebra, and not just a subspace, of $\d$ in order to define the corresponding Lie group $K$.

Proceeding as in the proof of Proposition \ref{prop: bc double}, we consider the smooth equivariant function $\widetilde{g} : \Sigma \times \CP \to G^\CC$ defined as follows. Let $\widetilde{g} \coloneqq \widehat{g}$ on the complement of $\Sigma \times (U_{x_+} \sqcup U_{x_-})$. Define $\widetilde{g}$ locally in small open discs $D^r_{x_\pm} \subset U_{x_\pm}$ around the points $x_\pm$ as $(\widetilde{g}|_{D^r_{x_+}}, \widetilde{g}|_{D^r_{x_-}}) \coloneqq a (\widehat{g}|_{x_+}, \widehat{g}|_{x_-})$ for any smooth $a : \Sigma \to K$ of our choice. Note here that $\widehat{g}|_{x_\pm} \in G$ by the equivariance of $\widehat{g}$ since $x_\pm \in \RR$.
We can then extend the definition of $\widetilde{g}$ to $\Sigma \times (U_{x_+} \sqcup U_{x_-})$ as we did in \S\ref{sec: bc double} so that $\widetilde{g}_{x_\pm} = \widetilde{g}|_{\Sigma \times U_{x_\pm}}$ depends only on $\sigma$, $\tau$ and the radial coordinate $|\xi_{x_\pm}|$ around $x_\pm$. In other words, $\widetilde{g}$ satisfies the archipelago conditions $(ii)$ and $(iii)$ on $U_{x_\pm}$.

It remains to show that $u = \widetilde{g} \widehat{g}^{-1}$, \emph{i.e.} the gauge transformation parameter from $\widehat{g}$ to $\widetilde{g}$, is equivariant and satisfies \eqref{allowed u simple}. The equivariance is clear from that of $\widetilde{g}$ and $\widehat{g}$. Now note that from the relation $\widetilde{g} = u \widehat{g}$ it follows that
\begin{equation} \label{u xpm eq 1}
(\widetilde{g}|_{x_+}, \widetilde{g}|_{x_-}) = (u|_{x_+}, u|_{x_-}) (\widehat{g}|_{x_+}, \widehat{g}|_{x_-}).
\end{equation}
But since $(\widetilde{g}|_{x_+}, \widetilde{g}|_{x_-}) = a (\widehat{g}|_{x_+}, \widehat{g}|_{x_-})$ we deduce that $(u|_{x_+}, u|_{x_-}) = a \in K$, which is the required condition \eqref{allowed u simple}.
\end{proof}
\end{proposition}

\subsection{Boundary conditions at complex conjugate simple poles} \label{sec: bc complex simple}

Let $x_\pm \in \bm z$ be simple poles of $\omega$ with $x_- = \overline{x_+}$, so that $\res_{x_-} \omega = \overline{\res_{x_+} \omega}$. By the equivariance property \eqref{equi omega A} of $A$ it also follows that $\tau(A_i|_{x_+}) = A_i|_{x_-}$ for $i = \tau, \sigma$.

The boundary equation of motion \eqref{bc simple pole general} can then be rewritten as
\begin{equation} \label{bc opposite res complex}
\epsilon_{ij} \langle\!\langle A_i|_{x_+}, \delta A_j|_{x_+} \rangle\!\rangle_{\g^\CC; x_\pm} = 0.
\end{equation}
Here $\langle\!\langle \cdot, \cdot \rangle\!\rangle_{\g^\CC; x_\pm} : \g^\CC \times \g^\CC \to \RR$ is the non-degenerate symmetric invariant bilinear form on the complexification $\g^\CC$, regarded as a real Lie algebra, defined by
\begin{equation*}
\langle\!\langle \ms x, \ms x' \rangle\!\rangle_{\g^\CC; x_\pm} \coloneqq 2 \Re \big( (\res_{x_+} \omega) \langle \ms x, \ms x' \rangle \big)
\end{equation*}
for any $\ms x, \ms x' \in \g^\CC$, where we denote by $\Re z$ and $\Im z$ the real and imaginary parts of a complex number $z$, respectively. When $\res_{x_+} \omega = - \res_{x_-} \omega$ so that $\res_{x_+} \omega \in \ii \RR$ this reduces, up to an overall factor, to the standard bilinear form $\Im \langle \ms x, \ms x' \rangle$ on $\g^\CC$.

The discussion below is completely analogous to that of \S\ref{sec: bc real simple}, just working with the complexification $\g^\CC$ rather than the real double $\d$. We will thus be much briefer in the arguments presented and only highlight the differences with \S\ref{sec: bc real simple}.

In particular, we can satisfy \eqref{bc opposite res complex} by choosing a Manin pair $(\g^\CC, \k)$, this time for the complexification rather than the real double, and demanding that
\begin{equation} \label{bc simple pole complex}
A_i|_{x_+} \in \k,
\end{equation}
for $i= \tau, \sigma$, noting that this implies $\delta A_i|_{x_+} \in \k$.

\medskip

Let $K$ denote the Lie subgroup of $G^\CC$ with Lie algebra $\k \subset \g^\CC$.

\begin{proposition} \label{prop: bc complex simple}
Suppose that $A$ satisfies the boundary condition \eqref{bc simple pole complex}, and we are given a field $\widehat{g}$ satisfying \eqref{A holo gauge 2} for which the archipelago condition $(i)$ holds.

Then the value of $\widehat{g}$ on the islands $U_{x_\pm}$ can be modified, without changing its value outside, so as to also satisfy the remaining archipelago conditions $(ii)$ and $(iii)$.

Furthermore, the value $g_{x_+} : \Sigma \to G^\CC$ of the archipelago type function $\widehat{g}$ at the point $x_+$ can be adjusted using $g_{x_+} \mapsto a g_{x_+}$ for any smooth function $a : \Sigma \to K$.
\begin{proof}
Evaluating \eqref{gauge transformation u} at $x_+$ yields $A^u|_{x_+} = - (d u u^{-1})|_{x_+} + u|_{x_+} A|_{x_+} u|_{x_+}^{-1}$. So a parameter $u$ such that
\begin{equation} \label{allowed u simple complex}
u|_{x_+} \in K
\end{equation}
defines an allowed gauge transformation.

We proceed as in the proof of Proposition \ref{prop: bc real simple} to construct a smooth equivariant $\widetilde{g} : \Sigma \times \CP \to G^\CC$ which is equal to $\widehat{g}$ on the complement of $\Sigma \times (U_{x_+} \sqcup U_{x_-})$ and which satisfies both of the archipelago conditions $(ii)$ and $(iii)$ on the islands $U_{x_\pm}$. Referring to the notation introduced in \S\ref{sec: bc real simple}, in the present case we let $\widetilde{g}|_{D^r_{x_+}} \coloneqq a \widehat{g}|_{x_+}$ for some smooth $a : \Sigma \to K$ of our choice. The rest of the definition of $\widehat{g}$ over $U_{x_+}$ is as in \S\ref{sec: bc real simple} and then we also let $\widetilde{g}|_{U_{x_-}} \coloneqq \tau(\widetilde{g}|_{U_{x_+}})$.

The fact that $u = \widehat{g} \widetilde{g}^{-1}$ is equivariant and satisfies \eqref{allowed u simple complex} is established as in \S\ref{sec: bc real simple} with minor changes. Specifically, we have
\begin{equation} \label{u xpm eq 1}
\widetilde{g}|_{x_+} = u|_{x_+} \widehat{g}|_{x_+}.
\end{equation}
But since $\widetilde{g}|_{x_+} = a \widehat{g}|_{x_+}$ we deduce that $u|_{x_+} = a \in K$, which is the condition \eqref{allowed u simple complex}, as required.
\end{proof}
\end{proposition}

\subsection{Manin triples and Poisson-Lie $T$-duality} \label{sec: Manin triple}

In all examples where $\omega$ has simple poles we shall be interested in the special case where the Manin pair $(\d,\k)$ (resp. $(\g^\CC, \k)$) can be extended to a Manin triple $(\d, \k, \p)$ (resp. $(\g^\CC, \k, \p)$). That is, $\p$ is another Lagrangian subalgebra of $\d$ (resp. $\g^\CC$) which is complementary to $\k$, \emph{i.e.} we have a direct sum $\d = \k \dotplus \p$ (resp. $\g^\CC = \k \dotplus \p$). We denote by $\dotplus$ the direct sum as vector spaces.

\medskip

An important class of Manin triples is given by a choice of solution $R \in \End \g$ of the modified classical Yang-Baxter equation
\begin{equation} \label{CYBE}
[R \ms x, R \ms y] - R \big( [R \ms x, \ms y] + [\ms x, R \ms y] \big) = - c^2 [\ms x, \ms y]
\end{equation}
for every $\ms x, \ms y \in \g$, where either $c = 1$ or $c = \ii$. We shall be particularly interested in solutions which are skew-symmetric with respect to the bilinear form $\langle \cdot, \cdot \rangle$ on $\g$, namely such that
\begin{equation*}
\langle R\ms x, \ms y \rangle = - \langle \ms x, R \ms y \rangle
\end{equation*}
for any $\ms x, \ms y \in \g$.

Specifically, in the real case where $c=1$ we define
\begin{equation*}
\g_R \coloneqq \{ ((R-1)\ms x, (R+1)\ms x) \,|\, \ms x \in \g \}, \qquad
\g^\delta \coloneqq \{ (\ms x, \ms x) \,|\, \ms x \in \g \}.
\end{equation*}
It is clear that $\g^\delta$ is a Lie subalgebra of $\d$, and it follows from \eqref{CYBE} that $\g_R$ also is. Suppose that $\d$ is equipped with its standard bilinear form, namely
\begin{equation*}
\langle\!\langle (\ms x, \ms y), (\ms x', \ms y') \rangle\!\rangle_\d \coloneqq \langle \ms x, \ms x' \rangle - \langle \ms y, \ms y' \rangle
\end{equation*}
for any $\ms x, \ms y, \ms x', \ms y' \in \g$. This corresponds, up to an overall factor, to the bilinear form considered in \S\ref{sec: bc real simple} when $\res_{x_-} \omega = - \res_{x_+} \omega$. In this case $\g^\delta$ is clearly isotropic and so is $\g_R$ by the skew-symmetry of $R$. It follows that $(\d, \g_R, \g^\delta)$ is a Manin triple.

In the complex case we take $c=\ii$ and define
\begin{equation*}
\g_R \coloneqq \{ (R-\ii) \ms x \,|\, \ms x \in \g \},
\end{equation*}
with $\g \subset \g^\CC$ denoting the real subalgebra of $\g^\CC$ regarded itself as a real Lie algebra. It follows again from \eqref{CYBE} that $\g_R$ is a Lie subalgebra of $\g^\CC$. Suppose, moreover, that $\g^\CC$ is equipped with its standard bilinear form, namely
\begin{equation*}
\langle\!\langle \ms x, \ms x' \rangle\!\rangle_{\g^\CC} = \Im \langle \ms x, \ms x' \rangle
\end{equation*}
for any $\ms x, \ms x' \in \g^\CC$, which corresponds to the bilinear form considered in \S\ref{sec: bc complex simple} with $\res_{x_-} \omega = - \res_{x_+} \omega$. In this case we have that $\g$ is certainly isotropic and $\g_R$ also is by the skew-symmetry of $R$. Therefore $(\g^\CC, \g_R, \g)$ is a Manin triple.

\medskip

Consider the Lie subgroup $G^\delta \coloneqq \{ (x, x) \,|\, x \in G \} \subset D$ with the Lie algebra $\g^\delta$. Also let $G_R$ denote the Lie subgroup of $D$ with Lie algebra $\g_R$. We will assume that the decomposition $\d = \g_R \dotplus \g^\delta$ lifts to the Lie group level, \emph{i.e.} that $D = G_R G^\delta$, or at least that $G_R G^\delta$ forms a dense subset of $D$. It then follows that a natural parametrisation of the quotient $G_R \backslash D$ in the case $c = 1$ is given by elements of $G^\delta$.

Likewise, in the case $c = \ii$, we let $G_R \subset G^\CC$ denote the Lie subgroup with Lie algebra $\g_R \subset \g^\CC$. Again, we will assume that the decomposition $\g^\CC = \g_R \dotplus \g$ similarly lifts to the Lie group level, \emph{i.e.} that $G^\CC = G_R G$, or at least that $G_R G$ forms a dense subset of $G^\CC$. A natural parametrisation of the quotient $G_R \backslash G^\CC$ is then given by elements of $G$. An example is provided by the Iwasawa decomposition $G^\CC = A N G$ where here $G$ is the compact real form of $G^\CC$ and $G_R = AN$.

\medskip

Since a Manin triple $(\d, \k, \p)$ (resp. $(\g^\CC, \k, \p)$) gives rise to two Manin pairs, namely $(\d, \k)$ or $(\d, \p)$ (resp. $(\g^\CC, \k)$ or $(\g^\CC, \p)$), we can apply the construction of \S\ref{sec: bc real simple} (resp. \S\ref{sec: bc complex simple}) at a pair of simple poles $x_\pm$ of $\omega$ using either of these Manin pairs. We expect the corresponding models obtained as in \S\ref{sec: 2d models} to be Poisson-Lie $T$-dual \cite{Klimcik:1995ux, Klimcik:1995dy}.

The main example of Poisson-Lie $T$-duality is provided by Manin triples of the form $\d = \g_R \dotplus \g^\delta$ or $\g^\CC = \g_R \dotplus \g$. This includes the Poisson-Lie $T$-duality between the Yang-Baxter $\sigma$-model, discussed in \S\ref{sec: YB}, and the $\lambda$-deformation of the principal chiral model, discussed in \S\ref{sec: lambda}. See for instance \cite{Vicedo:2015pna, Hoare:2015gda, Sfetsos:2015nya, Klimcik:2015gba}.

The Yang-Baxter $\sigma$-model with WZ-term, discussed in \S\ref{sec: YB-WZ} below, was also shown in \cite{Demulder:2017zhz} to be Poisson-Lie $T$-dual to itself for a different choice of parameters. In this case as well the duality is underpinned by certain choice of Manin triple so that it can also be described in the present formalism.

\medskip

Let us finally note that another way of ensuring the vanishing of the terms in the boundary equation of motion \eqref{bc explicit} corresponding to a pair of simple poles $x_\pm$ of $\omega$, is to ask that the terms associated with $x_+$ and with $x_-$ separately vanish. In other words, instead of \eqref{bc simple pole general} one could impose the weaker condition
\begin{equation} \label{bc simple pole CY}
\epsilon_{ij} \langle A_i|_{x_\pm}, \delta A_j|_{x_\pm} \rangle = 0.
\end{equation}
This situation was discussed in detail in \cite[\S9.1]{Costello:2017dso}. In particular, it was argued that \eqref{bc simple pole CY} can be satisfied by fixing a Manin triple $(\g, \mathfrak{l}_+, \mathfrak{l}_-)$, \emph{i.e.} making a choice of Lagrangian subalgebras $\mathfrak{l}_\pm \subset \g$ with $\g = \mathfrak{l}_+ \dotplus \mathfrak{l}_-$, and requiring that $A_i|_{x_\pm}$ be $\mathfrak{l}_\pm$-valued.
In the present language, working with such Manin triples on $\g$, as opposed to ones on $\g^\CC$ or $\d$, corresponds to considering skew-symmetric solutions $R \in \End \g$ of the modified classical Yang-Baxter equation \eqref{CYBE} for which $R^2 = 1$. The two subalgebras $\mathfrak{l}_\pm$ then correspond to the two eigenspaces $\ker(R\mp 1)$ of $R$.

\subsection{Generalised boundary conditions at a real double pole} \label{sec: bc double gen}

In light of the discussion of boundary conditions at pairs of simple poles in \S\ref{sec: bc real simple} and \S\ref{sec: bc complex simple}, we will now consider more general boundary conditions that can be imposed at double poles.
The algebraic setting of this section is similar to the one used in \cite{Klimcik:2018vhl} in the context of $\E$-models.

Let $x \in \bm z$ be a double pole of $\omega$ along the real axis, as in \S\ref{sec: bc double}. One can rewrite the boundary equation of motion \eqref{bc double pole general} in the following way. We consider the semi-direct product $\t \coloneqq \g \ltimes \g_{\rm ab}$ where $\g_{\rm ab}$ is an abelian copy of $\g$ on which $\g$ acts by the adjoint action. That is, $\t$ is isomorphic to the direct sum $\g \oplus \g$ as a vector space with Lie bracket given by $[(\ms x,\ms y), (\ms x',\ms y')]_\t = ([\ms x,\ms x'], [\ms x, \ms y'] - [\ms x', \ms y])$ for any $\ms x, \ms y, \ms x', \ms y' \in \g$. By the equivariance property of $A$ in \eqref{equi omega A}, since $x \in \RR$ we have $A_i|_x \in \g$. Also
\begin{equation*}
\tau \big( (\partial_{\xi_x} A_i)|_x \big) = \big( \tau (\partial_{\xi_x} A_i) \big) \big|_x
= \big( \partial_{\bar{\xi}_x} (\tau A_i) \big) \big|_x = \big( \mu_{\rm t}^\ast (\partial_{\xi_x} A_i) \big) \big|_x = (\partial_{\xi_x} A_i)|_x,
\end{equation*}
where the second step is by the anti-linearity of $\tau$, the third by the equivariance of $A$ and the last step follows because $\mu_{\rm t} x = x$. Hence $(\partial_{\xi_x} A_i)|_x \in \g$. We can therefore regard $(A_i|_x, (\partial_{\xi_x} A_i)|_x)$ as valued in $\t$, which allows us to rewrite \eqref{bc double pole general} as
\begin{equation} \label{double pole bc general}
\epsilon_{ij} \langle\!\langle (A_i|_x, (\partial_{\xi_x} A_i)|_x), \delta (A_j|_x, (\partial_{\xi_x} A_j)|_x) \rangle\!\rangle_{\t; x} = 0,
\end{equation}
where $\langle\!\langle \cdot, \cdot \rangle\!\rangle_{\t; x} : \t \times \t \to \RR$ is the bilinear form on $\t$ defined by
\begin{equation*}
\langle\!\langle (\ms x, \ms y), (\ms x', \ms y') \rangle\!\rangle_{\t; x} \coloneqq (\res_x \omega) \langle \ms x, \ms x' \rangle + (\res_x \xi_x \omega) \big( \langle \ms x, \ms y' \rangle + \langle \ms x', \ms y \rangle \big),
\end{equation*}
for every $\ms x, \ms y, \ms x', \ms y' \in \g$. One checks that this bilinear form is non-degenerate (using the fact that $\res_x \xi_x \omega \neq 0$ since $x$ is a double pole of $\omega$), symmetric and invariant.

The reformulation \eqref{double pole bc general} of the general condition \eqref{bc double pole general} leads to a natural way of imposing boundary conditions at the real double pole $x$, mimicking the discussion of \S\ref{sec: bc real simple} and \S\ref{sec: bc complex simple} for pairs of simple poles. Specifically, if we have a Manin pair $(\t, \k)$, \emph{i.e.} a Lagrangian subalgebra $\k$ of $\t$, then we can satisfy \eqref{double pole bc general} by requiring that
\begin{equation} \label{bc double pole gen}
(A_i|_x, (\partial_{\xi_x} A_i)|_x) \in \k
\end{equation}
for $i = \tau, \sigma$, noting that this then also implies $\delta (A_i|_x, (\partial_{\xi_x} A_i)|_x) \in \k$. For technical reasons to be discussed below, to do with making $\widehat{g}$ of archipelago type, we need to assume that the subalgebra $\g \ltimes \{ 0 \} \subset \t$ is complementary to our choice of Lagrangian subalgebra $\k \subset \t$. That is, we assume that we have a direct sum decomposition
\begin{equation} \label{complement in t}
\t = (\g \ltimes \{ 0 \}) \dotplus \k.
\end{equation}

Before proceeding, we note that the simple boundary condition \eqref{bc double pole} considered in \S\ref{sec: bc double} is a special case of \eqref{bc double pole gen}. Indeed, an obvious choice of Lagrangian subalgebra of $\t$ satisfying the condition \eqref{complement in t} is the abelian subalgebra $\{ 0 \} \ltimes \g_{\rm ab}$. Imposing the condition \eqref{bc double pole gen} in the case $\k = \{ 0 \} \ltimes \g_{\rm ab}$ is equivalent to requiring \eqref{bc double pole}.

\begin{proposition} \label{prop: bc double gen}
Suppose that $A$ satisfies the boundary condition \eqref{bc double pole gen}, and we are given a field $\widehat{g}$ satisfying \eqref{A holo gauge 2} for which the archipelago condition $(i)$ holds.

Then the value of $\widehat{g}$ on the island $U_x$ can be modified, without changing its value outside, so as to also satisfy the remaining archipelago conditions $(ii)$ and $(iii)$.
\begin{proof}
In order for $A^u$ to satisfy the second condition in \eqref{equi omega A} we should require that the function $u : \Sigma \times \CP \to G^\CC$ be equivariant, \emph{i.e.} $\tau u = \mu_{\rm t}^\ast u$. Evaluating the latter condition at the real pole $x$ implies that $u|_x \in G$ since $\mu_{\rm t} x = x$.
Also, we have
\begin{align*}
\tau \big( (\partial_{\xi_x} u)|_x u|_x^{-1} \big) &= \big( \tau (\partial_{\xi_x} u) \big) \big|_x (\tau u)|_x^{-1}
= \big( \partial_{\bar{\xi}_x} (\tau u) \big) \big|_x (\tau u)|_x^{-1}\\
&= \big( \mu_{\rm t}^\ast (\partial_{\xi_x} u) \big) \big|_x (\mu_{\rm t}^\ast u)|_x^{-1} = (\partial_{\xi_x} u)|_x u|_x^{-1},
\end{align*}
where in the second equality we use the anti-linearity of $\tau$ and in the third equality the equivariance of $u$. Therefore $(\partial_{\xi_x} u)|_x u|_x^{-1} \in \g$. We thus obtain a function
\begin{equation*}
U \coloneqq \big( u|_x, (\partial_{\xi_x} u)|_x u|_x^{-1} \big) : \Sigma \longrightarrow T
\end{equation*}
valued in the Lie group $T \coloneqq G \ltimes \g_{\rm ab}$ with Lie algebra $\t = \g \ltimes \g_{\rm ab}$.

Next, we determine conditions on $u$ for $A^u$ to still satisfy the boundary condition \eqref{bc double pole gen}. Evaluating \eqref{gauge transformation u} at $x$ we obtain
\begin{subequations}
\begin{equation} \label{Au double}
A_i^u|_x = - \partial_i (u|_x) u|_x^{-1} + u|_x A_i|_x u|_x^{-1}.
\end{equation}
On the other hand, differentiating \eqref{gauge transformation u} first with respect to the local holomorphic coordinate $\xi_x$ before evaluating at $x$ we find
\begin{align} \label{dAu double}
(\partial_{\xi_x} A_i^u)|_x &= - \partial_i \big( (\partial_{\xi_x} u)|_x u|_x^{-1} \big) + \big[ \partial_i (u|_x) u|_x^{-1}, (\partial_{\xi_x} u)|_x u|_x^{-1} \big] \notag\\
&\qquad\qquad + \big[ (\partial_{\xi_x} u)|_x u|_x^{-1}, u|_x A_i|_x u|_x^{-1} \big] + u|_x (\partial_{\xi_x}  A_i)|_x u|_x^{-1}.
\end{align}
\end{subequations}
Combining \eqref{Au double} and \eqref{dAu double} we thus find
\begin{equation} \label{Au bc preserve double}
\big( A_i^u|_x, (\partial_{\xi_x} A_i^u)|_x \big) = - \partial_i U U^{-1} + U \big( A_i|_x, (\partial_{\xi_x}  A_i)|_x \big) U^{-1},
\end{equation}
where the first term on the right hand side denotes the components of the Darboux derivative of $U : \Sigma \to T$ while the second term denotes the adjoint action of $U \in T$ on $\big( A_i|_x, (\partial_{\xi_x}  A_i)|_x \big) \in \t$. These are given explicitly by (see for instance \cite{Klimcik:2018vhl})
\begin{align*}
\partial_i (h, \ms v) (h, \ms v)^{-1} &= \big( \partial_i h h^{-1}, \partial_i \ms v - \big[ \partial_i h h^{-1}, \ms v \big] \big),\\
(k, \ms w) (\ms x, \ms y) (k, \ms w)^{-1} &= (k \ms x k^{-1}, k \ms y k^{-1} + [\ms w, \ms x]),
\end{align*}
for any smooth functions $h : \Sigma \to G$, $\ms v : \Sigma \to \g$ and any elements $k \in G$, $\ms w, \ms x, \ms y \in \g$.

It now follows from \eqref{Au bc preserve double} that an allowed gauge transformation, in the present case, should have parameter $u$ such that
\begin{equation} \label{allowed u double complex}
U = \big( u|_x, (\partial_{\xi_x} u)|_x u|_x^{-1} \big) \in K,
\end{equation}
where $K$ is the Lie subgroup of $T$ with Lie algebra $\k \subset \t$. We will assume that the decomposition \eqref{complement in t} lifts to a factorisation at the group level, namely that
\begin{equation} \label{complement in T}
T = K (G \ltimes \{ 0 \}).
\end{equation}

Having determined the set of allowed gauge transformations, we should find one which brings the smooth function $\widehat{g}$ to the desired archipelago form in $U_x$.

We proceed exactly as in the proofs of Propositions \ref{prop: bc real simple} and \ref{prop: bc complex simple}, by considering a smooth equivariant $\widetilde{g} : \Sigma \times \CP \to G^\CC$ defined as follows.
Let $\widetilde{g} \coloneqq \widehat{g}$ on the complement of $\Sigma \times U_x$. We then define $\widetilde{g}$ as being constant in a small open disc $D^r_x \subset U_x$ around $x$ by letting $(\widetilde{g}|_{D^r_x}, 0) \in G \ltimes \{ 0 \}$ be the representative of the class  in $K \backslash T$ of $(\widehat{g}|_x, (\partial_{\xi_x} \widehat{g})|_x \widehat{g}|_x^{-1}) \in T$. Note that here we have made use of the property \eqref{complement in T}. The reason we had to choose a representative in $G \ltimes \{ 0 \}$ is that we want $\widetilde{g}$ to be of archipelago type on the island $U_x$, which by definition means $\widetilde{g}|_{D^r_x}$ is constant along $\CP$ so that necessarily $(\partial_{\xi_x} \widetilde{g} \widetilde{g}^{-1})|_{D^r_x} = 0$.
Finally, we can also extend the definition of $\widetilde{g}$ to $\Sigma \times U_x$ as we did in \S\ref{sec: bc double} so that $\widetilde{g}_x = \widetilde{g}|_{\Sigma \times U_x}$ depends only on $\sigma$, $\tau$ and the radial coordinate $|\xi_x|$ around $x$. Therefore, by construction $\widetilde{g}$ satisfies both of the archipelago conditions $(ii)$ and $(iii)$ on $U_x$.

Now consider the gauge transformation parameter $u = \widetilde{g} \widehat{g}^{-1}$. Its equivariance is clear from that of $\widetilde{g}$ and $\widehat{g}$. And from the relation $\widetilde{g} = u \widehat{g}$ we obtain
\begin{equation*}
\widetilde{g}|_x = u|_x \widehat{g}|_x, \qquad
0 = (\partial_{\xi_x} u)|_x u|_x^{-1} + u|_x (\partial_{\xi_x} \widehat{g})|_x \widehat{g}|_x^{-1} u|_x^{-1}.
\end{equation*}
The second equality is obtained by computing $\partial_{\xi_x} \widetilde{g} \widetilde{g}^{-1}$ in terms of $u$ and $\widehat{g}$ and then evaluating at $x$, noting that since $\widetilde{g}$ is constant along $\CP$ in a neighbourhood of $x$ we have $(\partial_{\xi_x} \widetilde{g} \widetilde{g}^{-1})|_x = 0$.
By definition of the product in $T = G \ltimes \g_{\rm ab}$ the above two equations are equivalent to
\begin{equation*}
( \widetilde{g}|_x, 0) = \big( u|_x, (\partial_{\xi_x} u)|_x u|_x^{-1} \big) \big( \widehat{g}|_x, (\partial_{\xi_x} \widehat{g})|_x \widehat{g}|_x^{-1} \big).
\end{equation*}
Yet since $( \widetilde{g}|_x, 0)$ was defined as the representative in $G \ltimes \{ 0 \}$ of the class in $K \backslash T$ of $(\widehat{g}|_x, (\partial_{\xi_x} \widehat{g})|_x \widehat{g}|_x^{-1}) \in T$, the condition \eqref{allowed u double complex} follows.
\end{proof}
\end{proposition}

An important class of Lie subalgebras $\k \subset \t$ with the property \eqref{complement in t} is provided by solutions $R \in \End \g$ of the classical Yang-Baxter equation, \emph{i.e.} \eqref{CYBE} with $c = 0$, which reads
\begin{equation} \label{CYBE0}
[R \ms x, R \ms y] - R \big( [R \ms x, \ms y] + [\ms x, R \ms y] \big) = 0
\end{equation}
for every $\ms x, \ms y \in \g$. Specifically, given such a solution we define the Lie subalgebra
\begin{equation*}
\g_R \coloneqq \{ (- R\ms x, \ms x) \,|\, \ms x \in \g \}
\end{equation*}
of $\t$. The fact that it is a subalgebra is a direct consequence of \eqref{CYBE0}. Indeed, for any $\ms x, \ms y \in \g$ we have
\begin{equation*}
\big[ (-R\ms x, \ms x), (-R\ms y, \ms y) \big]_\t = \big( [- R\ms x, - R \ms y], [- R\ms x, \ms y] - [- R\ms y, \ms x] \big) = (- R \ms z, \ms z) \in \g_R
\end{equation*}
where $\ms z = - [R\ms x, \ms y] - [\ms x, R \ms y] \in \g$.

In the case when $\res_x \omega = 0$, which we shall focus on in \S\ref{sec: hYB}, it is clear that the Lie subalgebra $\g \ltimes \{ 0 \} \subset \t$ is isotropic with respect to $\langle\!\langle \cdot, \cdot \rangle\!\rangle_{\t; x}$. If, moreover, the solution $R \in \End \g$ of \eqref{CYBE0} is skew-symmetric in the sense that
\begin{equation*}
\langle R\ms x, \ms y \rangle = - \langle \ms x, R \ms y \rangle
\end{equation*}
for any $\ms x, \ms y \in \g$, then the subalgebra $\g_R \subset \t$ is also isotropic. In this case we therefore have a Manin triple $(\t, \g_R, \g \ltimes \{ 0 \})$.

\medskip

Let $G_R$ denote the Lie subgroup of $T$ with Lie algebra $\g_R$. We will assume, as in \eqref{complement in T}, that the vector space direct sum decomposition $\t = (\g \ltimes \{0\}) \dotplus \g_R$ lifts to the Lie group level, namely that $T = G_R (G \ltimes \{ 0 \})$, or at least that $G_R (G \ltimes \{ 0 \})$ forms a dense subset of $T$, cf. \S\ref{sec: Manin triple}.

\section{Examples} \label{sec: examples}

In this section we rederive the actions of many known integrable $\sigma$-models from the four-dimensional Chern-Simons action \eqref{holo CS action}. Specifically, our starting point in each case is the 1-form $\omega$ given by
\begin{equation*}
\omega = \varphi(z) dz
\end{equation*}
where $\varphi(z)$ is the twist function of the integrable $\sigma$-model that we want to consider, which has at most double poles. We then impose natural boundary conditions on the 1-form $A$ at the poles of $\omega$, of the various types discussed in \S\ref{sec: boundary conditions}. In each case we then compute the corresponding action \eqref{action sigma model} and show that it coincides with the known action of the given integrable $\sigma$-model. In all cases, we also find that the meromorphic 1-form $\mathcal L$ coincides with the Lax connection of the integrable $\sigma$-model.

In every example, $\omega$ will have a pair of simple zeroes, say at $y_\pm \in \bm \zeta$. Since all the $\sigma$-models that we want to reconstruct are relativistic, by Remark \ref{rem: relativistic} we will thus take $\sigma_{y_\pm} = \sigma^\pm$ in the notation of \eqref{L pole structure}. The reason for not taking $\sigma_{y_\pm}$ both equal to $\sigma^+$ or both equal to $\sigma^-$ is that the resulting 1-form $\mathcal L$ would be quite degenerate, with one of its light-cone components being independent of the spectral parameter. In the absence of a Lax connection, there is no guarantee that the resulting $\sigma$-model would be integrable. We will come back in \S\ref{sec: E-model} to considering such a case.

\subsection{Principal chiral model with WZ-term} \label{sec: PCM+WZ}

Although the action for this model was already derived from \eqref{holo CS action} in \cite{Costello:2019tri}, we give the derivation of this case in detail as it illustrates the general procedure for constructing the action of an integrable $\sigma$-model from the two-dimensional action \eqref{action sigma model} in the simplest possible setting.

\medskip

Consider the 1-form (see for instance \cite[\S5.1.3)]{Vicedo:2017cge} and \cite{Maillet:1985ec} in the case $k=0$)
\begin{equation*}
\omega = K \frac{1 - z^2}{(z-k)^2} dz,
\end{equation*}
where $K$ and $k$ are real parameters. It has a pair of double poles at $k \in \RR$ and $\infty$. 
Note that, under the change of variable $z \mapsto z + k$, this can also be brought to the equivalent form
\begin{equation*}
\omega = - K \frac{(z - z_+)(z - z_-)}{z^2} dz
\end{equation*}
with $z_\pm \coloneqq -k\pm 1$. This is the 1-form used in \cite{Costello:2019tri} to describe the principal chiral model with WZ-term.

As discussed in \S\ref{sec: bc double}, we can satisfy the boundary equations of motion \eqref{bc explicit} by requiring that
\begin{equation} \label{PCM bc}
A_i|_k = 0, \qquad
A_i|_\infty = 0,
\end{equation}
for $i = \tau, \sigma$. It follows from Lemma \ref{lem: bc} and Proposition \ref{prop: bc double} that $\widehat{g}$ can be chosen of archipelago type and, moreover, such that
\begin{equation*}
g_k = g, \qquad
g_\infty = 1
\end{equation*}
for some $g : \Sigma \to G$. The latter condition is used to fixed the gauge invariance of Proposition \ref{prop: gauge-inv 2d action}. Evaluating \eqref{A holo gauge} at $k$ and $\infty$ we then find
\begin{equation}\label{PCM bc bis}
A|_k = - dg g^{-1} + \Ad_g \mathcal L|_k, \qquad
A|_\infty = \mathcal L|_\infty.
\end{equation}

Now the 1-form $\omega$ has simple zeroes at $\pm 1$, \emph{i.e.} $\bm \zeta = \{1, -1\}$. On the other hand, we also know from combining the second equations in \eqref{PCM bc} and \eqref{PCM bc bis} that $\mathcal L$ vanishes at infinity. Thus $U_\sigma = U_\tau = 0$ in the general expression \eqref{L pole structure} for the meromorphic dependence of $\mathcal L$ on $\CP$. As discussed at the start of this section, in the general notation of \eqref{L pole structure} we choose $\sigma_{\pm 1} = \sigma^\pm$, so that the Lax connection in the present case takes the form
\begin{equation*}
\mathcal L = \frac{V^{1}}{z-1} d\sigma^+ + \frac{V^{-1}}{z+1} d\sigma^-,
\end{equation*}
for some $V^{\pm 1} : \Sigma \to \g$. Their expressions in terms of the $G$-valued field $g$ can now be determined uniquely by solving $- \partial_i g g^{-1} + \Ad_g \mathcal L_i|_k = 0$ for $i = \tau, \sigma$, which follows from combining the first two equations in \eqref{PCM bc} and \eqref{PCM bc bis}. We find
\begin{equation*}
V^{\pm 1} = (k\mp 1) j_\pm,
\end{equation*}
where $j_\pm \coloneqq g^{-1} \partial_\pm g$.

\medskip

We now have all the ingredients to compute the action \eqref{action sigma model} in the case at hand. Note that the terms in this action corresponding to the pole $\infty \in \bm z$ do not contribute since we chose to set $g_\infty = 1$. To compute the first term we thus only need the residue
\begin{equation*}
\res_k \omega\wedge \L =-K \big( (k-1) j_+ d\sigma^++(k+1) j_- d\sigma^- \big),
\end{equation*}
while for the WZ-term we note that $\res_k\omega=-2K k$.
From these expressions and the fact that $d\sigma^+ \wedge d\sigma^- = \ha d\sigma \wedge d\tau$ we finally obtain
\begin{align*}
S[g] &= \frac{K}{2}\int\langle j_+, j_- \rangle d\sigma \wedge d\tau + K k \, I_{\rm WZ}[g],
\end{align*}
which we recognize as the action of the principal chiral model in the presence of a WZ-term.

\subsection{Homogeneous Yang-Baxter $\sigma$-model} \label{sec: hYB}

We will follow the conventions of \cite[\S 4.2.1]{Delduc:2019bcl}.

The procedure for constructing a homogeneous Yang-Baxter deformation \cite{Kawaguchi:2014qwa} of a given integrable $\sigma$-model does not modify the underlying twist function \cite{Vicedo:2015pna}. For this reason, we start from the same 1-form as in \S\ref{sec: PCM+WZ}. However, to simplify the discussion we set $k = 0$ and take
\begin{equation*}
\omega = K \frac{1 - z^2}{z^2} dz
\end{equation*}
where $K$ is a real parameter. The discussion of the more general case with $k \neq 0$ could be done by proceeding along the same lines as in \S\ref{sec: YB-WZ} below.

Now although the 1-form $\omega$ is the same as in \S\ref{sec: PCM+WZ}, here we will impose a different boundary condition at its double pole $0$ compared to that in \eqref{PCM bc}.
More precisely, we will replace it with a boundary condition that is associated with a choice of Lagrangian subalgebra of $\t = \g \ltimes \g_{\rm ab}$, as discussed in \S\ref{sec: bc double gen}.

\medskip

The bilinear form on $\t$ in the present case reads
\begin{equation*}
\langle\!\langle (\ms x, \ms y), (\ms x', \ms y') \rangle\!\rangle_{\t; 0} = K \big( \langle \ms x, \ms y' \rangle + \langle \ms x', \ms y \rangle \big).
\end{equation*}
Let us fix any skew-symmetric solution of the classical Yang-Baxter equation \eqref{CYBE0}. As recalled at the end of \S\ref{sec: bc double gen}, it follows that $\g_R = \{ (- R\ms x, \ms x) \,|\, \ms x \in \g \}$ is a Lagrangian Lie subalgebra of $\t$.

We may use this Lagrangian subalgebra $\g_R \subset \t$ to satisfy the boundary equations of motion \eqref{bc explicit} by requiring that (see \S\ref{sec: bc double gen})
\begin{equation} \label{hYB bc}
\big( A_i|_0, (\partial_z A_i)|_0 \big) \in \g_R, \qquad
A_i|_\infty = 0,
\end{equation}
for $i = \tau, \sigma$. Recall here that $\xi_0 = z$ is the local coordinate at $0$. By virtue of Lemma \ref{lem: bc} and Proposition \ref{prop: bc double gen} we can choose $\widehat{g}$ to be of archipelago type and, moreover, such that
\begin{equation*}
g_0 = g, \qquad
g_\infty = 1
\end{equation*}
for some $g : \Sigma \to G$. The latter condition fixes the gauge invariance of Proposition \ref{prop: gauge-inv 2d action}. Evaluating \eqref{A holo gauge} at $0$ and $\infty$ we then find
\begin{subequations} \label{hYB bc bis}
\begin{equation} \label{hYB bc bis 1}
A|_0 = - dg g^{-1} + \Ad_g \mathcal L|_0, \qquad
A|_\infty = \mathcal L|_\infty,
\end{equation}
but also, taking the derivative with respect to $z$ before evaluating at $0$ and using the fact that $(\partial_z \widehat{g})|_0 = 0$ by virtue of the archipelago condition $(iii)$, we obtain
\begin{equation} \label{hYB bc bis 2}
(\partial_z A)|_0 = \Ad_g (\partial_z \mathcal L)|_0.
\end{equation}
\end{subequations}

Since $\mathcal L$ is meromorphic with poles in the set $\bm \zeta = \{ 1, - 1 \}$ of zeroes of $\omega$ and since it vanishes at infinity by the last two equations in \eqref{hYB bc} and \eqref{hYB bc bis 1}, it follows from the general expression \eqref{L pole structure} that we can write
\begin{equation*}
\mathcal L = \frac{V^{1}}{z - 1} d\sigma^+ + \frac{V^{-1}}{z + 1} d\sigma^-.
\end{equation*}
Now the first condition in \eqref{hYB bc} implies that $A_i|_0 = - R (\partial_z A_i)|_0$. By combining this with \eqref{hYB bc bis} and the above explicit form of $\mathcal L$ we obtain
\begin{equation*}
V^{\pm 1} = \mp \frac{1}{1 \pm R_g} j_\pm,
\end{equation*}
where $R_g \coloneqq \Ad_{g^{-1}} \circ R \circ \Ad_g$.

Finally, noting that $\res_0 \omega \wedge \mathcal L = - K (V^1 d\sigma^+ + V^{-1} d\sigma^-)$ and $\res_0 \omega = 0$ we find that the action \eqref{action sigma model} reduces to
\begin{align*}
S[g] &= \frac{K}{2} \int_{\Sigma} \bigg\langle j_+, \frac{1}{1 - R_g} j_- \bigg\rangle d\sigma \wedge d\tau.
\end{align*}
This is the action of the homogeneous Yang-Baxter deformation of the principal chiral model, as first constructed in \cite{Kawaguchi:2014qwa} in the case of the semi-symmetric space $\sigma$-model.

\subsection{Yang-Baxter $\sigma$-model} \label{sec: YB}

The twist function in this case was first computed in \cite{Delduc:2013fga}. We will follow the conventions of \cite[\S 4.2.2]{Delduc:2019bcl}. In particular, we take
\begin{equation*}
\omega = \frac{K}{1-c^2 \eta^2} \frac{1 - z^2}{z^2 - c^2 \eta^2} dz,
\end{equation*}
with $K, \eta$ real parameters and $c = 1$ or $c = \ii$.

We fix a skew-symmetric solution $R \in \End \g$ of the modified classical Yang-Baxter equation \eqref{CYBE}. As $\res_{c \eta} \omega = - \res_{- c \eta} \omega$, it follows from \S\ref{sec: Manin triple} that $\g_R$ is a Lagrangian subalgebra of $\d$ when $c = 1$ (resp. $\g^\CC$ when $c = \ii$). The boundary equations of motion \eqref{bc explicit} can then be satisfied by requiring that
\begin{subequations} \label{YB bc}
\begin{equation}
(A_i|_{\eta}, A_i|_{-\eta}) \in \g_R, \qquad
A_i|_\infty = 0,
\end{equation}
for $i = \tau, \sigma$, in the case $c = 1$, or
\begin{equation}
A_i|_{\ii \eta} \in \g_R, \qquad
A_i|_\infty = 0,
\end{equation}
\end{subequations}
for $i = \tau, \sigma$, in the case $c = \ii$. It follows from Lemma \ref{lem: bc} and Propositions \ref{prop: bc double}, \ref{prop: bc real simple} and \ref{prop: bc complex simple} that we can choose $\widehat{g}$ to be of archipelago type.

Moreover, by the discussion in \S\ref{sec: Manin triple} and Proposition \ref{prop: gauge-inv 2d action}, we are able to choose our archipelago type field $\widehat{g}$ such that
\begin{equation*}
g_{\pm c \eta} = g, \qquad
g_\infty = 1
\end{equation*}
for some $g : \Sigma \to G$. More precisely, by the last part of Proposition \ref{prop: bc real simple} (resp. Proposition \ref{prop: bc complex simple}), the value of $\widehat{g}$ at the pair of points $\pm \eta$ when $c = 1$ (resp. at the point $\ii \eta$ when $c = \ii$) defines a field on $\Sigma$ valued in $G_R \backslash D$ (resp. in $G_R \backslash G^\CC$). We can parametrise this quotient by the diagonal subgroup $G^\delta$ (resp. the real subgroup $G$) which allows us to choose $\widehat{g}$ such that $(\widehat{g}|_\eta, \widehat{g}|_{-\eta}) = (g, g)$ (resp. $\widehat{g}|_{\ii \eta} = g$).
In the case $c = \ii$ we then use the fact that $\widehat{g}$ is equivariant to obtain also $g_{-\ii \eta} = \tau(g_{\ii \eta}) = g$.
With this choice, evaluating \eqref{A holo gauge} at the poles of $\omega$ we then obtain
\begin{equation} \label{YB A from L}
A|_{\pm c \eta} = - dg g^{-1} + \Ad_g \mathcal L|_{\pm c \eta}, \qquad
A|_\infty = \mathcal L|_\infty.
\end{equation}

Now $\omega$ has simple zeroes at $\pm 1$ so $\bm \zeta = \{ 1, -1 \}$. Moreover, combining the last two equations in \eqref{YB bc} and \eqref{YB A from L} we find that $\mathcal L$ should vanish at infinity. By the same reasoning as in \S\ref{sec: PCM+WZ}, this allows us to write the Lax matrix in the form
\begin{equation*}
\mathcal L = \frac{V^1}{z-1} d\sigma^+ + \frac{V^{-1}}{z+1} d\sigma^-
\end{equation*}
for some $\g$-valued fields $V^{\pm 1}$ to be determined.

It follows from the first condition in \eqref{YB bc} that $(R+c) A_i|_{c \eta} = (R - c) A_i|_{-c \eta}$. By combining this with the first equation in \eqref{YB A from L} and the above explicit rational form of $\mathcal L$ we therefore deduce that
\begin{align*}
&- (R+c) dg g^{-1} + (R+c) \Ad_g \bigg( \frac{1}{c\eta-1} V^1 d\sigma^+ + \frac{1}{c\eta+1} V^{-1} d\sigma^- \bigg),\\
&\qquad = - (R-c) dg g^{-1} - (R-c) \Ad_g \bigg( \frac{1}{c\eta+1} V^1 d\sigma^+ + \frac{1}{c\eta-1} V^{-1} d\sigma^- \bigg).
\end{align*}
By equating the $d\sigma^\pm$-components on both sides we obtain two equations for the two unknowns $V^{\pm 1}$ which can be solved to give
\begin{equation*}
V^{\pm 1} = \pm \frac{c^2\eta^2-1}{1 \pm \eta R_g} j_\pm
\end{equation*}
where $R_g = \Ad_{g^{-1}} \circ R \circ \Ad_g$ as before and $j_\pm = g^{-1} \partial_\pm g$.

\medskip

Since $g_\infty = 1$ there is no WZ-term in the action \eqref{action sigma model} corresponding to the double pole at $\infty$. On the other hand, as $\res_{\pm c \eta} \omega = \pm K/2 c \eta$ and $g_{c \eta} = g_{-c\eta}$, it follows that the WZ-terms associated with the simple poles $\pm c \eta$ cancel out.

To compute the first term in the action \eqref{action sigma model} we need the residue
\begin{equation*}
\res_{\pm c \eta} \omega \wedge \mathcal L = (\res_{\pm c \eta} \omega) \mathcal L|_{\pm c \eta} = \pm \frac{K}{2 c \eta} \bigg( \frac{c\eta+1}{1 \pm \eta R_g} j_\pm d\sigma^\pm - \frac{c\eta-1}{1 \mp \eta R_g} j_\mp d\sigma^\mp \bigg).
\end{equation*}
Putting everything together we find that the action \eqref{action sigma model} becomes
\begin{equation*}
S[g] = \frac{K}{2} \int_{\Sigma} \bigg\langle j_+, \frac{1}{1 - \eta R_g} j_- \bigg\rangle d\sigma \wedge d\tau
\end{equation*}
which coincides with the Yang-Baxter $\sigma$-model action \cite{Klimcik:2002zj, Klimcik:2008eq}.

\subsection{$\lambda$-deformation of the principal chiral model} \label{sec: lambda}

The twist function in this case was first computed in \cite{Hollowood:2014rla}.
We shall follow here the conventions of \cite[\S 4.4]{Delduc:2019bcl}. In particular, we take
\begin{equation*}
\omega = \frac{K}{1-\alpha^2} \frac{1 - z^2}{z^2 - \alpha^2} dz, \qquad \lambda = \frac{1+\alpha}{1-\alpha},
\end{equation*}
with $K, \alpha$ real parameters.

Since $\res_{\alpha} \omega = - \res_{- \alpha} \omega$, it follows from \S\ref{sec: Manin triple} that $\g^\delta$ is a Lagrangian subalgebra of $\d$. We can therefore satisfy the boundary condition \eqref{bc explicit} by requiring that
\begin{equation*}
(A_i|_{\alpha}, A_i|_{-\alpha}) \in \g^\delta, \qquad A_i|_\infty = 0
\end{equation*}
for $i= \tau, \sigma$. In other words, we have $A_i|_{\alpha} = A_i|_{-\alpha}$ and $A_i|_\infty = 0$.
It follows from Lemma \ref{lem: bc} and Propositions \ref{prop: bc double} and \ref{prop: bc real simple} that we can choose $\widehat{g}$ to be of archipelago type. Now as in the corresponding discussion of \S\ref{sec: YB}, it follows from the last part of Proposition \ref{prop: bc real simple} that $(\widehat{g}|_\alpha, \widehat{g}|_{-\alpha})$ defines a field on $\Sigma$ valued in $G^\delta \backslash D$. A natural parametrisation of this quotient consists of elements of the form $(h, 1)$ for $h \in G$. We can thus choose our archipelago type field $\widehat{g}$ such that
\begin{equation*}
g_\alpha = g, \qquad
g_{-\alpha} = 1, \qquad
g_\infty = 1
\end{equation*}
for some $g : \Sigma \to G$. The condition on $g_\infty$ is imposed by virtue of Proposition \ref{prop: gauge-inv 2d action}. Evaluating \eqref{A holo gauge} at the poles of $\omega$ we thus obtain
\begin{equation} \label{A from L lambda}
A|_\alpha = - dg g^{-1} + \Ad_g \mathcal L|_\alpha, \qquad
A|_{-\alpha} = \mathcal L|_{-\alpha}, \qquad
A|_\infty = \mathcal L|_\infty.
\end{equation}

Using the last equation and the boundary condition at infinity we get $\mathcal L|_\infty = 0$. Since $\mathcal L$ is meromorphic with simple poles in the set $\bm \zeta = \{ 1, - 1 \}$ of zeroes of $\omega$ we deduce its dependence on $z$ to be of the form, cf. \S\ref{sec: PCM+WZ}, \S\ref{sec: hYB} and \S\ref{sec: YB},
\begin{equation*}
\mathcal L = \frac{\alpha+1}{z-1} U_+ d\sigma^+ + \frac{\alpha+1}{z+1} U_- d\sigma^-
\end{equation*}
for some $\g$-valued pair of fields $U_\pm = (\alpha + 1)^{-1} V^{\pm 1}$ on $\Sigma$. The normalising factor of $\alpha+1$ is introduced for convenience. In particular, evaluating $\mathcal L$ at $\pm \alpha$ we find
\begin{equation} \label{L pm alpha}
\mathcal L|_\alpha = -\lambda U_+ d\sigma^+ + U_- d\sigma^-, \qquad
\mathcal L|_{-\alpha} = - U_+ d\sigma^+ + \lambda U_- d\sigma^-.
\end{equation}
It then follows from the boundary conditions at $\pm \alpha$ and the first two equations in \eqref{A from L lambda} that
\begin{equation*}
- dg g^{-1} - \lambda \Ad_g U_+ d\sigma^+ + \Ad_g U_- d\sigma^- = - U_+ d\sigma^+ + \lambda U_- d\sigma^-
\end{equation*}
Equating the coefficients of $d\sigma^\pm$ on both sides, solving for $U_\pm$ and substituting back into \eqref{L pm alpha} we find
\begin{equation*}
\mathcal L|_\alpha = - \frac{\lambda \Ad_g}{1 - \lambda \Ad_g} j_+ d\sigma^+ + \frac{\Ad_g}{\Ad_g - \lambda} j_- d\sigma^-.
\end{equation*}
We did not specify $\mathcal L|_{-\alpha}$ since it will not be needed as $g_{-\alpha} = 1$. It now follows that
\begin{equation*}
\res_\alpha \omega \wedge \L = (\res_\alpha \omega) \mathcal L|_\alpha = 2k \frac{\lambda \Ad_g}{1 - \lambda \Ad_g} j_+ d\sigma^+ - 2k \frac{\Ad_g}{\Ad_g - \lambda} j_- d\sigma^-
\end{equation*}
using the fact that $\res_\alpha \omega = - 2k$ where $k = - K/4\alpha$.

\medskip

Inserting all of the above into the action \eqref{action sigma model} we find it simplifies to
\begin{align*}
S[g] &= \frac{k}{2} \int_\Sigma \langle g^{-1} \partial_+ g, g^{-1} \partial_- g \rangle d\sigma \wedge d\tau + k \, I_{\rm WZ}[g]\\
&\qquad\qquad + k \int_\Sigma \bigg\langle \frac{1}{\lambda^{-1} - \Ad_g} \partial_+ g g^{-1}, g^{-1} \partial_- g \bigg\rangle d\sigma \wedge d\tau.
\end{align*}
It coincides with the action of the $\lambda$-deformation of the principal chiral model \cite{Sfetsos:2013wia}, written using the conventions of \cite[\S 4.4]{Delduc:2019bcl}.

\subsection{Bi-Yang-Baxter $\sigma$-model} \label{sec: biYB}

We follow the conventions used in \cite{Delduc:2015xdm}. In particular, we take
\begin{equation} \label{omegaBYB}
\omega = \frac{16 K z}{\zeta^2 (z - z_+)(z - z_-)(z - \tilde z_+)(z - \tilde z_-)} dz,
\end{equation}
where $K \in \RR$. The four poles $z_\pm$ and $\tilde z_\pm$ as well as $\zeta \in \RR$ are related to the two real deformation parameters $\eta$ and $\tilde \eta$ of the model by
\begin{gather*}
z_\pm = \frac{- 2 \rho \pm \ii \eta}{\zeta}, \qquad
\tilde z_\pm = - \frac{2 + 2 \rho \pm \ii \tilde \eta}{\zeta}, \qquad
\rho = - \ha \bigg( 1 - \frac{\eta^2 - \tilde \eta^2}{4} \bigg), \\
\zeta^2 = \bigg(1 + \frac{(\eta + \tilde \eta)^2}{4} \bigg) \bigg(1 + \frac{(\eta - \tilde \eta)^2}{4} \bigg).
\end{gather*}

Choose two skew-symmetric solutions $R, \tilde R \in \End \g$ of the modified Yang-Baxter equation \eqref{CYBE} with $c = \ii$. Because $\res_{z_-} \omega = - \res_{z_+} \omega$ and $\res_{\tilde z_-} \omega = - \res_{\tilde z_+} \omega$ it follows from \S\ref{sec: Manin triple} that $\g_R$ and $\g_{\tilde R}$ are both Lagrangian subalgebras of $\g^\CC$. To satisfy the boundary equations of motion \eqref{bc explicit} we impose that
\begin{equation} \label{bYB bc}
A_i|_{z_+} \in \g_R, \qquad
A_i|_{\tilde z_+} \in \g_{\tilde R},
\end{equation}
for $i = \tau, \sigma$.
By Lemma \ref{lem: bc} and Propositions \ref{prop: bc complex simple} we can choose $\widehat{g}$ to be of archipelago type. And by the discussion in \S\ref{sec: Manin triple}, see also the corresponding discussion in \S\ref{sec: YB}, we can take $\widehat{g}$ such that
\begin{equation} \label{g biYB}
g_{z_\pm} = g, \qquad
g_{\tilde z_\pm} = \tilde g
\end{equation}
for some $g, \tilde g : \Sigma \to G$.
Evaluating \eqref{A holo gauge} at the poles of $\omega$ we obtain
\begin{equation} \label{biYB A from L}
A|_{z_\pm} = - dg g^{-1} + \Ad_g \mathcal L|_{z_\pm}, \qquad
A|_{\tilde z_\pm} = - d\tilde g \tilde g^{-1} + \Ad_{\tilde g} \mathcal L|_{\tilde z_\pm}.
\end{equation}

The 1-form $\omega$ has a simple zero at the origin and at infinity, that is $\bm \zeta = \{ 0, \infty \}$. In the present case, the general form \eqref{L pole structure} of the Lax connection therefore reads
\begin{equation} \label{biYB Lax}
\mathcal L = \bigg( B_+ + \frac{\zeta}{2} z J_+ \bigg) d\sigma^+ + \bigg( B_- + \frac{\zeta}{2} z^{-1} J_- \bigg) d\sigma^-
\end{equation}
for some $\g$-valued fields $B_\pm \coloneqq U_\tau \pm U_\sigma$, $J_+ \coloneqq 2 \zeta^{-1} V^\infty$ and $J_- \coloneqq 2 \zeta^{-1} V^0$ to be determined.

The $d\sigma^\pm$-components of the two equations
\begin{equation*}
(R+\ii)A_i|_{z_+} = (R-\ii)A_i|_{z_-}, \qquad (\tilde R+\ii)A_i|_{\tilde z_+} = (\tilde R-\ii)A_i|_{\tilde z_-},
\end{equation*}
which follow from \eqref{bYB bc}, give us four equations on the four unknowns $B_\pm$ and $J_\pm$. Explicitly, we have
\begin{equation} \label{jpm jtpm eq}
j_\pm = B_\pm \pm \frac{\eta}{2} R_g J_\pm - \rho J_\pm, \qquad \tilde \jmath_\pm = B_\pm \mp \frac{\tilde \eta}{2} \tilde R_{\tilde g} J_\pm - (\rho+1) J_\pm,
\end{equation}
where we have introduced $j_\pm \coloneqq g^{-1} \partial_\pm g$ and $\tilde \jmath_\pm \coloneqq \tilde g^{-1} \partial_\pm \tilde g$.
Taking the difference of these two equations yields
\begin{equation*}
J_\pm = \frac{1}{1 \pm \frac{\eta}{2} R_g \pm \frac{\tilde \eta}{2} \tilde R_{\tilde g}} (j_\pm - \tilde \jmath_\pm).
\end{equation*}
The first equation in \eqref{jpm jtpm eq} then also yields $B_\pm = j_\pm \mp \frac{\eta}{2} R_g J_\pm + \rho J_\pm$. In particular, the Lax connection \eqref{biYB Lax} thus coincides with \cite[(3.4.9)]{Lacroix:2018njs} or, up to a conventional sign, with \cite[(2.18)]{Delduc:2015xdm}.

\medskip

We have $\res_{z_\pm} \omega = \mp \frac{2 \ii K}{\eta}$ and $\res_{\tilde z_\pm} \omega = \mp \frac{2 \ii K}{\tilde \eta}$. It then follows from \eqref{g biYB} that the four WZ-terms in the action \eqref{action sigma model} cancel in pairs. We also have
\begin{align*}
\res_{z_+} \omega \wedge \mathcal L + \res_{z_-} \omega \wedge \mathcal L &= 2 K (J_+ d\sigma^+ - J_- d\sigma^-),\\
\res_{\tilde z_+} \omega \wedge \mathcal L + \res_{\tilde z_-} \omega \wedge \mathcal L &= - 2 K (J_+ d\sigma^+ - J_- d\sigma^-)
\end{align*}
so that the action \eqref{action sigma model} takes the final form
\begin{align*}
S[g, \tilde g] = K \int_{\Sigma} \langle j_+ - \tilde \jmath_+, J_- \rangle d\sigma \wedge d\tau.
\end{align*}
This is the action of the bi-Yang-Baxter $\sigma$-model as written in \cite[(2.2)]{Delduc:2015xdm}.

Note that, contrary to the examples discussed in all the previous sections, as well as in \S\ref{sec: YB-WZ} below, we have not fixed the gauge invariance of Proposition \ref{prop: gauge-inv 2d action} by fixing the value of $\widehat{g}$ at any of the poles of $\omega$. It follows that the above action still has the gauge invariance of Proposition \ref{prop: gauge-inv 2d action} which here takes the form
\begin{equation*}
(g, \tilde g) \longmapsto (g h, \tilde g h)
\end{equation*}
for any smooth $h : \Sigma \to G$. Fixing this gauge invariance by setting $\tilde g = 1$ we obtain the original action of the bi-Yang-Baxter $\sigma$-model \cite{Klimcik:2008eq, Klimcik:2014bta}.

\subsection{Yang-Baxter $\sigma$-model with WZ-term} \label{sec: YB-WZ}

Consider the 1-form \cite{Delduc:2014uaa}
\begin{equation*}
\omega = \frac{K(1 - z^2)}{(z - k)^2 - c^2 \mathcal A^2} dz,
\end{equation*}
with free parameters $K, k, \mathcal A \in \RR$. We shall consider in parallel the cases when $c = 1$ and $c = \ii$. Note that in the limit $k \to 0$ we recover the 1-form of the Yang-Baxter $\sigma$-model with $\mathcal A = \eta$, discussed in \S\ref{sec: YB}, up to an overall factor.

Besides the double pole at $\infty$, the 1-form $\omega$ has two simple poles at $z_\pm = k \pm c \mathcal A$ which are both real for $c=1$ and complex conjugate for $c = \ii$.
However, in order to apply the construction of \S\ref{sec: bc real simple} (resp. \S\ref{sec: bc complex simple}) to the pair of simple poles $z_\pm$, we require a Lagrangian subalgebra of $\d$ (resp. $\g^\CC$).
But since the residues
\begin{equation*}
\res_{z_\pm} \omega = \pm K \frac{1 - z_\pm^2}{2c\mathcal A}
\end{equation*}
are such that $\res_{z_-} \omega \neq - \res_{z_+} \omega$, the bilinear form $\langle \!\langle \cdot, \cdot \rangle\!\rangle_{\d; z_\pm}$ on $\d$ (resp. $\langle \!\langle \cdot, \cdot \rangle\!\rangle_{\g^\CC; z_\pm}$ on $\g^\CC$) is not the standard one, by contrast with the situations of \S\ref{sec: YB}, \S\ref{sec: lambda} and \S\ref{sec: biYB}.
Our analysis, at least in the case $c=\ii$, is closely related to that of \cite{Klimcik:2017ken} where the double $\g^\CC$ is also equipped with the more general bilinear form $\langle \!\langle \cdot, \cdot \rangle\!\rangle_{\g^\CC; z_\pm}$.

A consequence of the bilinear form on $\d$ (resp. $\g^\CC$) not being the standard one is that the diagonal subalgebra $\g^\delta \subset \d$ (resp. the real subalgebra $\g \subset \g^\CC$) is no longer isotropic. Moreover, given any skew-symmetric solution $R \in \End \g$ of the modified classical Yang-Baxter equation \eqref{CYBE}, the corresponding subalgebra $\g_R$ of $\d$ (resp. of $\g^\CC$) will in general not be isotropic either.

\medskip

To construct a Lagrangian subalgebra of $\d$ (resp. of $\g^\CC$) we proceed as follows.
Let $R \in \End \g$ be a skew-symmetric solution of \eqref{CYBE} such that
\begin{equation} \label{R2 eq}
R^3 = c^2 R.
\end{equation}
This implies that $R$ is diagonalisable with $\g = \g_+ \dotplus \g_0 \dotplus \g_-$ its eigenspace decomposition where $\g_\pm \coloneqq \ker(R \mp c)$ and $\g_0 \coloneqq \ker R$ are subalgebras of $\g$, and moreover that $[\g_0, \g_\pm] \subset \g_\pm$ and $\g_0$ is abelian (see for instance \cite[Proposition C.2.2]{Lacroix:2018njs}). In particular, we can thus write $R = c (\pi_+ - \pi_-)$ where $\pi_\pm$ and $\pi_0$ are the projections onto the subalgebras $\g_\pm$ and $\g_0$ relative to the eigenspace decomposition of $R$.

It is useful to note that $\pi_0 = - c^2 R^2 + 1$ which is symmetric with respect to the bilinear form $\langle \cdot, \cdot \rangle$ on $\g$. The relation \eqref{R2 eq} then implies that $\pi_0 R = R \pi_0 = 0$. Let
\begin{equation} \label{R tilde non-skew}
\tilde R \coloneqq R + \theta \pi_0 \in \End \g
\end{equation}
for some real parameter $\theta \in \RR$ to be fixed shortly.

Since $\pi_0 \in \End \g$ is symmetric, it follows that $\tilde R$ is not skew-symmetric. However, one checks that it still satisfies the modified classical Yang-Baxter equation \eqref{CYBE}, for the same value of $c$ (see for instance \cite[Theorem C.2.1]{Lacroix:2018njs}). So $\g_{\tilde R}$ defined as in \S\ref{sec: Manin triple} is a subalgebra of $\d$ complementary to the diagonal subalgebra $\g^\delta$ if $c=1$ or a subalgebra of $\g^\CC$ complementary to the real subalgebra $\g$ if $c = \ii$.

Moreover, we find that $\g_{\tilde R}$ is isotropic, and so in fact Lagrangian, with respect to the bilinear form $\langle \!\langle \cdot, \cdot \rangle\!\rangle_{\d; z_\pm}$ on $\d$ (resp. $\langle \!\langle \cdot, \cdot \rangle\!\rangle_{\g^\CC; z_\pm}$ on $\g^\CC$) provided that
\begin{equation*}
(\theta - c)^2 (\res_{z_+} \omega) + (\theta + c)^2 (\res_{z_-} \omega) = 0.
\end{equation*}
Of the two solutions for $\theta \in \RR$, the one which is regular in the limit $k \to 0$ reads
\begin{equation*}
\theta = \frac{- c^2 k \eta^2}{(1 - c^2 \eta^2)\mathcal A},
\end{equation*}
where the real parameter $\eta$ is related to the parameters $\mathcal A$ and $k$ as (see \cite{Kawaguchi:2011mz, Kawaguchi:2013gma, Delduc:2014uaa} in the case $c = \ii$)
\begin{equation*}
\mathcal A = \eta \sqrt{1 - \frac{k^2}{1 - c^2 \eta^2}}.
\end{equation*}
It therefore follows that $\g_{\tilde R}$, with $\tilde R \in \End \g$ defined in \eqref{R tilde non-skew} and for $\theta \in \RR$ as above, is a Lagrangian subalgebra of $\d$ (resp. of $\g^\CC$). In other words, we have a Manin pair $(\d, \g_{\tilde R})$ (resp. $(\g^\CC, \g_{\tilde R})$), which we can use in the construction of \S\ref{sec: bc real simple} (resp. \S\ref{sec: bc complex simple}).

\medskip

Concretely, we will realise the boundary equations of motion \eqref{bc explicit} by demanding that
\begin{subequations} \label{YB-WZ bc}
\begin{equation}
(A_i|_{z_+}, A_i|_{z_-}) \in \g_{\tilde R}, \qquad
A_i|_\infty = 0,
\end{equation}
for $i =\tau, \sigma$, in the case $c = 1$, or
\begin{equation}
A_i|_{z_+} \in \g_{\tilde R}, \qquad
A_i|_\infty = 0,
\end{equation}
\end{subequations}
for $i = \tau, \sigma$, in the case $c = \ii$.
By virtue of Lemma \ref{lem: bc} and Propositions \ref{prop: bc double}, \ref{prop: bc real simple} and \ref{prop: bc complex simple} we can choose $\widehat{g}$ to be of archipelago type. Moreover, by the discussion in \S\ref{sec: Manin triple} and Proposition \ref{prop: gauge-inv 2d action}, we can take $\widehat{g}$ to be such that
\begin{equation} \label{g YB+WZ}
g_{z_\pm} = g, \qquad
g_\infty = 1
\end{equation}
for some $g : \Sigma \to G$. We refer to the corresponding discussion in \S\ref{sec: YB} for details. In the case $c = \ii$ we used here the equivariance of $\widehat{g}$ to show that $g_{z_-} = \tau(g_{z_+}) = g$.

Evaluating \eqref{A holo gauge} at the poles of $\omega$ we obtain
\begin{equation} \label{YB-WZ A from L}
A|_{z_\pm} = - dg g^{-1} + \Ad_g \mathcal L|_{z_\pm}, \qquad
A|_\infty = \mathcal L|_\infty.
\end{equation}
Combining the last two equations of \eqref{YB-WZ bc} and \eqref{YB-WZ A from L} we deduce that $\mathcal L$ vanishes at $\infty$. And since $\bm \zeta = \{ 1, -1 \}$ we can write the general form \eqref{L pole structure} as
\begin{equation*}
\mathcal L = \frac{V^1}{z-1} d\sigma^+ + \frac{V^{-1}}{z+1} d\sigma^-,
\end{equation*}
for some $V^{\pm 1} : \Sigma \to \g$ to be determined. Now it follows from the first condition in \eqref{YB-WZ bc} that $(\tilde R + c) A_i|_{z_+} = (\tilde R - c) A_i|_{z_-}$. We therefore obtain the two equations
\begin{equation*}
(\tilde R_g + c) \bigg( \! - j_\pm + \frac{V^{\pm 1}}{z_+ \mp 1} \bigg) = (\tilde R_g - c) \bigg( \! - j_\pm + \frac{V^{\pm 1}}{z_- \mp 1} \bigg),
\end{equation*}
for the two unknowns $V^{\pm 1}$, or in other words
\begin{equation*}
\bigg( \frac{c(z_+ + z_- \mp 2)}{(z_+ \mp 1)(z_- \mp 1)} + \frac{z_- - z_+}{(z_+ \mp 1)(z_- \mp 1)} \tilde R_g \bigg) V^{\pm 1} = 2 c j_\pm.
\end{equation*}
The operator on the left hand side can be inverted by making use of the relations $\pi_0 R = R \pi_0 = 0$, $\pi_0^2 = \pi_0$ and $R^2 = c^2(1 - \pi_0)$. We find
\begin{equation*}
V^{\pm 1} = \mp (1 \mp  k - c^2 \eta^2 \mp \mathcal A R_g + \eta^2 R^2_g) j_\pm.
\end{equation*}

By contrast with the situation of \S\ref{sec: YB}, the WZ-terms associated with the poles $z_\pm$ in the action \eqref{action sigma model} do not cancel since $\res_{z_+} \omega + \res_{z_-} \omega = - 2 K k$, which is non-zero.
On the other hand, we have
\begin{equation*}
\res_{z_+} \omega \wedge \mathcal L + \res_{z_-} \omega \wedge \mathcal L = - K (V^1 d\sigma^+ + V^{-1} d\sigma^-),
\end{equation*}
so that the action \eqref{action sigma model} evaluates to
\begin{align*}
S[g] &= \frac{K}{2} \int_{\Sigma} \big\langle j_-, \big(1 - c^2 \eta^2 - \mathcal A R_g + \eta^2 R^2_g\big) j_+ \big\rangle d\sigma \wedge d\tau + K k \, I_{\rm WZ}[g].
\end{align*}
This coincides, in the case when $c = \ii$, with the action of the Yang-Baxter $\sigma$-model with WZ-term as given in \cite[(2.7)]{Delduc:2014uaa}.

\section{$\E$-models} \label{sec: E-model}

We will take the 1-form $\omega$ to be given by
\begin{equation*}
\omega = K \frac{1 - z^2}{(z - z_+)(z - z_-)} dz
\end{equation*}
where, for simplicity, we restrict attention to the case $z_\pm \in \RR$. The reasoning below can be easily adapted to the case of complex conjugate simple poles.

Even though the starting point $\omega$ is of the same form as in \S\ref{sec: YB}, \S\ref{sec: lambda} and \S\ref{sec: YB-WZ}, and each of the integrable $\sigma$-models considered in those sections are known to be examples of $\E$-models, see \cite{Klimcik:2015gba} and \cite{Klimcik:2017ken} respectively, we will proceed very differently to construct the underlying $\E$-models themselves.

\medskip

To begin with, we will impose a very different boundary condition on the 1-form $A$, at the poles $z_\pm$ of $\omega$, to those considered in \S\ref{sec: YB}, \S\ref{sec: lambda} and \S\ref{sec: YB-WZ}. In fact, our choice of boundary condition is very closely related to that considered in \cite{Severa:2016prq} for deriving the $\E$-model from three-dimensional Chern-Simons theory.

Moreover, the choice of coordinates $\sigma_{\pm 1}$ that we will make for the pair of zeroes $\pm 1$ of $\omega$ in the general expression \eqref{L pole structure} for $\mathcal L$ will be different from that used throughout \S\ref{sec: examples}. Indeed, the choice will result in the $d\sigma$-component of the 1-form $\mathcal L$ being trivial.

\subsection{Boundary condition}

Evaluating $A$ at the pair of points $z_\pm$ yields a $\d$-valued 1-form $\mathsf A \coloneqq (A|_{z_+}, A|_{z_-})$ on $\Sigma$, whose components we denote
\begin{equation*}
\mathsf A_i \coloneqq (A_i|_{z_+}, A_i|_{z_-}) : \Sigma \longrightarrow \d
\end{equation*}
for $i = \tau, \sigma$. In terms of these, we can express the boundary equations of motion \eqref{bc explicit} as
\begin{equation} \label{bc E-model gen}
\langle\!\langle \mathsf A_\sigma, \delta \mathsf A_\tau \rangle\!\rangle_{\d; z_\pm} - \langle\!\langle \mathsf A_\tau, \delta \mathsf A_\sigma \rangle\!\rangle_{\d; z_\pm} = 0,
\end{equation}
where $\langle\!\langle \cdot, \cdot \rangle\!\rangle_{\d; z_\pm} : \d \times \d \to \RR$ is defined in \S\ref{sec: bc real simple}.

\medskip

Let $\E : \d \to \d$ be a linear map such that $\E^2 = \id$ which is symmetric with respect to the bilinear form $\langle\!\langle \cdot, \cdot \rangle\!\rangle_{\d; z_\pm}$ on $\d$. We shall impose the boundary conditions at the poles $z_\pm$ and $\infty$ of $\omega$ to be
\begin{equation} \label{E bc}
\mathsf A_\tau = \E(\mathsf A_\sigma), \qquad A_i|_\infty = 0,
\end{equation}
for $i = \tau, \sigma$. The first boundary condition provides a simple way of satisfying the boundary equation of motion \eqref{bc E-model gen}, merely as a consequence of the symmetry of the linear map $\E$.

Under a gauge transformation with parameter $u : \Sigma \times \CP \to G^\CC$, the components $\mathsf A_i$ for $i = \tau, \sigma$ of the $\d$-valued 1-form $\mathsf A$ become
\begin{equation*}
\mathsf A_i^{\mathsf u} \coloneqq (A^u_i|_{z_+}, A^u_i|_{z_-}) = - \partial_i \mathsf u \mathsf u^{-1} + \mathsf u \mathsf A_i \mathsf u^{-1},
\end{equation*}
where $\mathsf u \coloneqq (u|_{z_+}, u|_{z_-}) : \Sigma \to D$.
If we require that $\mathsf u = 1 \in D$ then the components $\mathsf A_i^{\mathsf u}$ for $i = \tau, \sigma$ of the gauge transformed connection $\mathsf A^{\mathsf u}$ are trivially seen to satisfy the boundary condition \eqref{E bc}. Therefore, any gauge transformation with parameter $u : \Sigma \times \CP \to G^\CC$ such that $u|_{z_\pm} = 1$ is allowed. Using such a gauge transformation one can then ensure that $\widehat{g} : \Sigma \times \CP \to G^\CC$ satisfying \eqref{A holo gauge 2} is of archipelago type, by the same arguments as in \S\ref{sec: bc} and \S\ref{sec: bc double}. Without loss of generality we can also choose the radii of the discs $U_{z_\pm}$ around the points $z_\pm$ to be equal, namely $R_{z_+} = R_{z_-}$. We shall denote this common radius by $R$.

As usual, we use Proposition \ref{prop: gauge-inv 2d action} to set $g_\infty = 1$. Then $\mathcal L_i|_\infty = A_i|_\infty = 0$, where the last step uses the second boundary condition in \eqref{E bc}. We take $\mathcal L$ of the form
\begin{equation} \label{L for E-model}
\mathcal L = \bigg( \frac{V^1}{z - 1} + \frac{V^{-1}}{z + 1} \bigg) d\tau
\end{equation}
where we have chosen $\sigma_1 = \sigma_{-1} = \tau$ in the notation of the general form \eqref{L pole structure}.

\subsection{Action}

Since $\mathcal L$ is regular at the simple poles $z_\pm \in \bm z$ of $\omega$ we have
\begin{equation*}
\res_{z_\pm} \omega \wedge \mathcal L = (\res_{z_\pm} \omega) \mathcal L|_{z_\pm}.
\end{equation*}
Let $\mathsf J_\tau \coloneqq (\mathcal L_\tau|_{z_+}, \mathcal L_\tau|_{z_-}) : \Sigma \to \d$ and
\begin{equation*}
\ell \coloneqq (g_{z_+}, g_{z_-}) : \Sigma \longrightarrow D.
\end{equation*}
We also let $\widehat{\ell} \coloneqq (\widehat{g}_{z_+}, \widehat{g}_{z_-}) : \Sigma \times [0, R] \to D$.
The action \eqref{action sigma model} can be rewritten in the present case as
\begin{align*}
S[\ell] = - \frac{1}{2} \int_{\Sigma} \langle\!\langle d \ell \ell^{-1}, \Ad_\ell \mathsf J_\tau d\tau \rangle\!\rangle_{\d; z_\pm} - \frac{1}{6} \int_{\Sigma \times [0, R]} \langle\!\langle d \widehat{\ell} \widehat{\ell}^{-1}, d \widehat{\ell} \widehat{\ell}^{-1} \wedge d \widehat{\ell} \widehat{\ell}^{-1} \rangle\!\rangle_{\d; z_\pm}.
\end{align*}
Evaluating \eqref{A holo gauge} at the pair of points $z_\pm$ yields
\begin{equation*}
\mathsf A_\tau = - \partial_\tau \ell \ell^{-1} + \ell \mathsf J_\tau \ell^{-1}, \qquad \mathsf A_\sigma = - \partial_\sigma \ell \ell^{-1}.
\end{equation*}
By combining this with the first boundary condition in \eqref{E bc} it follows that
\begin{equation} \label{Jtau Emodel}
\Ad_\ell \mathsf J_\tau = \partial_\tau \ell \ell^{-1} - \E(\partial_\sigma \ell \ell^{-1}).
\end{equation}
Finally, substituting this into the above action yields
\begin{align*}
S[\ell] &= - \frac{1}{2} \int_{\Sigma} \langle\!\langle \partial_\sigma \ell \ell^{-1}, \partial_\tau \ell \ell^{-1} \rangle\!\rangle_{\d; z_\pm} d\sigma \wedge d\tau + \frac{1}{2} \int_{\Sigma} \langle\!\langle \partial_\sigma \ell \ell^{-1}, \E (\partial_\sigma \ell \ell^{-1}) \rangle\!\rangle_{\d; z_\pm} d\sigma \wedge d\tau\\
&\qquad\qquad\qquad - \frac{1}{6} \int_{\Sigma \times [0, R]} \langle\!\langle d \widehat{\ell} \widehat{\ell}^{-1}, d \widehat{\ell} \widehat{\ell}^{-1} \wedge d \widehat{\ell} \widehat{\ell}^{-1} \rangle\!\rangle_{\d; z_\pm}.
\end{align*}
This is the action of the $\E$-model \cite{Klimcik:1995dy, Klimcik:1996nq} in the case of the real double $\d = \g \oplus \g$.

Note that since we have set $\mathcal L_\sigma = 0$, it follows from the equations of motion \eqref{eom1} expressed in terms of $\mathcal L$ that we have $\partial_\sigma \mathsf J_\tau = 0$. If $\Sigma = \RR^2$ and we assume that $\mathsf J_\tau$ vanishes at spatial infinity, then it follows that $\mathsf J_\tau = 0$. By virtue of \eqref{Jtau Emodel} this implies the on-shell relation $\partial_\tau \ell \ell^{-1} = \E(\partial_\sigma \ell \ell^{-1})$, which we also recognise as the equation of motion of the $\E$-model.

\section{Conclusion}

\subsection{Integrable coupled $\sigma$-models}

A general procedure for coupling together an arbitrary number of integrable $\sigma$-models in a way that preserves integrability was proposed in \cite{Delduc:2019bcl}. In particular, the action for an integrable $\sigma$-model coupling together $N$ copies of the principal chiral model with WZ-term, as given in \cite{Delduc:2018hty}, was constructed by first devising its Hamiltonian as that of an affine Gaudin model and then performing its inverse Legendre transform.

This same action was recently rederived in \cite{Costello:2019tri} starting from the four-dimensional action \eqref{holo CS action}. In fact, it follows from the results of the present paper that the action for this integrable $\sigma$-model can also be obtained directly from the two-dimensional action \eqref{2d action intro} by substituting for $\varphi$ and $\mathcal L$ the twist function and the Lax connection, respectively, of the affine Gaudin model constructed in \cite{Delduc:2019bcl} or in \cite{Lacroix:2019xeh} for the version with gauge invariance.

\subsection{$\lambda$-deformations and `doubled' Chern-Simons}

An appealing feature of the two-dimensional action \eqref{2d action intro} is its `universality'.

The $\lambda$-deformation, considered in \S\ref{sec: lambda}, is a particular example of an integrable $\sigma$-model that describes a certain integrable deformation. This was constructed for the principal chiral model in \cite{Sfetsos:2013wia}, for the symmetric and semi-symmetric space $\sigma$-models in \cite{Hollowood:2014rla, Hollowood:2014qma} and more recently for the pure-spinor superstring on the $AdS_5 \times S^5$ background in \cite{Benitez:2019oaw}.

It was, in fact, already known that the actions of $\lambda$-deformations can be written in the `universal' form \eqref{2d action intro}, see \cite[(3.98)]{Schmidtt:2018hop}. Explicitly, in the case of the $\lambda$-deformation of the principal chiral model it follows from \S\ref{sec: lambda} that the action reads
\begin{equation*}
S_\lambda[g] = k \int_\Sigma \langle g^{-1} dg, \mathcal L|_\alpha \rangle + k\, I_{\rm WZ}[g].
\end{equation*}
Interestingly, this action was obtained in \cite{Schmidtt:2017ngw, Schmidtt:2018hop} by starting from that of a `double' Chern-Simons theory, whose Lagrangian is given by a difference $CS(A_+) - CS(A_-)$ of two Chern-Simons 3-forms for $\g$-valued 1-forms $A_\pm$ on $D \times \RR$ where $D$ is a disc.

It would be interesting to derive this `double' Chern-Simons theory starting from the four-dimensional theory of \cite{Costello:2019tri}. More generally, one may wonder whether such an intermediate three-dimensional Chern-Simons theory also exists more generally for other integrable $\sigma$-models whose action takes the universal form \eqref{2d action intro}.

\medskip

Finally, in connection with the derivation of $\E$-models presented in \S\ref{sec: E-model}, it would also be interesting to understand the relationship between the approach of \cite{Costello:2019tri}, which we have been using, and the formalism of \cite{Severa:2016prq} in which $\E$-models can equally be obtained but by starting instead from three-dimensional Chern-Simons theory.

\subsection{Yang-Baxter type deformations}

In \S\ref{sec: YB}, \S\ref{sec: biYB} and \S\ref{sec: YB-WZ} we imposed boundary conditions at each pair of simple poles $x_\pm$ of $\omega$ by applying the general procedure outlined in \S\ref{sec: bc real simple} or \S\ref{sec: bc complex simple} with a choice of Lagrangian subalgebra $\k$ of the form $\g_R$ for some solution $R \in \End \g$ of the modified classical Yang-Baxter equation. This characterises the class of Yang-Baxter type deformations of integrable $\sigma$-models, obtained by splitting a double pole $x$ in $\omega$ into two simple poles $x_\pm$ as in \cite{Delduc:2013fga, Vicedo:2015pna, Delduc:2019bcl}. Within this class, there is a WZ-term associated with the pair of poles $x_\pm$ if and only if $\res_{x_+} \omega + \res_{x_-} \omega \neq 0$.

Just as \S\ref{sec: YB-WZ} generalises \S\ref{sec: YB} by introducing a WZ-term, one could also consider a similar generalisation of the construction of \S\ref{sec: biYB} by starting from a more general 1-form $\omega$ with two arbitrary pairs of simple poles $z_\pm$ and $\tilde z_\pm$ (respecting the reality conditions) but with $\res_{z_+} \omega + \res_{z_-} \omega \neq 0$ and $\res_{\tilde z_+} \omega + \res_{\tilde z_-} \omega \neq 0$. It is natural to conjecture that this would result in the bi-Yang-Baxter $\sigma$-model with WZ-term introduced in~\cite{Delduc:2017fib}. We leave the verification of this conjecture for future work.

\medskip

In fact, the 1-form $\omega$ considered in \cite[(14.2)]{Costello:2019tri} is precisely of the general type described above (up to a M\"obius transformation). The boundary conditions imposed on $A$ at each pair of simple poles $z_\pm$ and $\tilde{z}_\pm$ of $\omega$ in \cite[\S 14]{Costello:2019tri} are associated with a choice of Manin triple $(\tilde{\g},\mathfrak{l}_+, \mathfrak{l}_-)$ for the Lie algebra $\tilde{\g}=\g \oplus \tilde{\mathfrak{h}}$, where $\tilde{\mathfrak{h}}$ denotes an auxiliary copy of the Cartan subalgebra of $\g$ (considering $\tilde{\g}$ instead of $\g$ allows for a more direct construction of a Manin triple).

As recalled at the end of \S\ref{sec: Manin triple}, such boundary conditions correspond in the present language to choosing an $R$-matrix satisfying $R^2=1$. In light of the above discussion, this suggests that the trigonometric deformation of the principal chiral model constructed in~\cite[\S 14]{Costello:2019tri} coincides with a certain bi-Yang-Baxter $\sigma$-model with WZ-term on the Lie group $\tilde{G}$ corresponding to $\tilde \g$. We note, however, that its description in~\cite{Costello:2019tri} uses a different parametrisation of the degrees of freedom than the one that we used in \S \ref{sec: biYB} to describe the bi-Yang-Baxter $\sigma$-model. It would be interesting to make this relation more explicit.

\subsection{Boundary conditions \emph{vs} Representations}

The non-abelian $T$-dual of the principal chiral model was described in \cite{Delduc:2019bcl} as an affine Gaudin model. Indeed, a particular realisation of the relevant affine Gaudin model, with $\omega$ as in \S\ref{sec: hYB}, was shown to reproduce the Hamiltonian, phase space and action of this $\sigma$-model.

We have not considered this particular model here, part of the reason being that we do not expect the two-dimensional action \eqref{2d action intro} to hold in this case. Indeed, since $\res_0 \omega = 0$, a natural choice of Lagrangian subalgebra $\k \subset \t$ is given by $\k = \g \ltimes \{0\}$. The quotient $K \backslash T$ in this case is naturally parameterised by elements of the abelian subgroup $\{ 0 \} \ltimes \g_{\rm ab} \subset T$. In other words, the field of the resulting integrable $\sigma$-model would be $\g$-valued, so it is natural to conjecture that this corresponds to the non-abelian $T$-dual of the principal chiral model. However, the Lagrangian subalgebra $\k$ does not satisfy condition \eqref{complement in t} which was necessary to bring $\widehat{g}$ to the archipelago form in the proof of Proposition \ref{prop: bc double gen}. More precisely, the choice of representative in $\{ 0 \} \ltimes \g_{\rm ab}$ for the quotient $(G \ltimes \{ 0 \}) \backslash T$ is not compatible with the archipelago condition $(iii)$. This is why the action \eqref{2d action intro} does not hold. It would be interesting to see how the present construction can be generalised to this case.

\medskip

More generally, it would be very interesting to understand the precise connection between the choice of boundary condition on $A$ in the setting of \cite{Costello:2019tri} and the choice of realisation of a suitable infinite-dimensional Lie algebra in the setting of \cite{Vicedo:2017cge}. Recall, for instance, that affine Toda field theories were shown to admit an affine Gaudin model realisation in \cite{Vicedo:2017cge}. It would be important to clarify what boundary conditions, if any, can be imposed on $A$ in order to derive the action of affine Toda field theories from the four-dimensional action \eqref{holo CS action}.

With this in mind, it would also be very interesting to see whether the approach of \cite{Costello:2019tri} can be used to furnish new classes of models, for instance by identifying new types of suitable boundary conditions to be imposed on $A$. It would then also be interesting to understand what the corresponding infinite-dimensional Lie algebra representation is in the affine Gaudin model language.

\subsection{Quantising integrable $\sigma$-models}

Perhaps most importantly, the results of the present paper bring further evidence in support of the connection established in \cite{Vicedo:2019dej} between the two formalisms of \cite{Costello:2019tri} and \cite{Vicedo:2017cge}.

And while at present they have mainly been used to describe \emph{classical} integrable $\sigma$-models, one important interest in these new general frameworks lies in their potential in addressing the long-standing open problem of quantising integrable $\sigma$-models from first principles. We expect that exploiting the close connection between these two formalisms will be vital in making progress on this important question.

\end{document}